\documentclass[useAMS,usenatbib]{mnras}
\usepackage{newtxtext,newtxmath}
\usepackage[T1]{fontenc}
\DeclareRobustCommand{\VAN}[3]{#2}
\let\VANthebibliography\thebibliography
\def\thebibliography{\DeclareRobustCommand{\VAN}[3]{##3}\VANthebibliography}
\usepackage{graphicx}
\usepackage{subfig}
\usepackage{subcaption}
\usepackage{amsmath}	
\usepackage{fancyhdr}
\usepackage{hyperref}
\usepackage{todonotes}
\usepackage{academicons}
\usepackage{orcidlink}
\usepackage{xcolor}
\usepackage{soul}
\sethlcolor{yellow}
\usepackage{widetext}

\title[Precision Cosmology and Testing GR using GWs]{Prospect of Precision Cosmology and Testing General Relativity using Binary Black Holes- Galaxies Cross-correlation}

\author[Afroz and Mukherjee]{
  Samsuzzaman Afroz $^{1}$\orcidlink{0009-0004-4459-2981} \thanks{samsuzzaman.afroz@tifr.res.in},
  Suvodip Mukherjee $^{1}$\orcidlink{0000-0002-3373-5236} \thanks{suvodip@tifr.res.in} 
  \\
  $^{1}$Department of Astronomy and Astrophysics, Tata Institute of Fundamental Research, Mumbai 400005, India
}

\date{11 September 2024}
\pubyear{2024}

\begin{document}
\label{firstpage}
\pagerange{\pageref{firstpage}--\pageref{lastpage}}
\maketitle

\begin{abstract}

Modified theories of gravity predict deviations from General Relativity (GR) in the propagation of gravitational waves (GW) across cosmological distances. A key prediction is that the GW luminosity distance will vary with redshift, differing from the electromagnetic (EM) luminosity distance due to varying effective Planck mass. We introduce a model-independent, data-driven approach to explore these deviations using multi-messenger observations of dark standard sirens (Binary Black Holes, BBH). By combining GW luminosity distance measurements from dark sirens with Baryon Acoustic Oscillation (BAO) measurements, BBH redshifts inferred from cross-correlation with spectroscopic or photometric galaxy surveys, and sound horizon measurements from the Cosmic Microwave Background (CMB), we can make a data-driven test of GR (jointly with the Hubble constant) as a function of redshift. Using the multi-messenger technique with the spectroscopic DESI galaxy survey, we achieve precise measurements of deviations in the effective Planck mass variation with redshift. For the Cosmic Explorer and Einstein Telescope (CEET), the best precision is approximately 3.6\%, and for LIGO-Virgo-KAGRA (LVK), it is 7.4\% at a redshift of $\rm{z = 0.425}$. Additionally, we can measure the Hubble constant with a precision of about 1.1\% from CEET and 7\% from LVK over five years of observation with a 75\% duty cycle. We also explore the potential of cross-correlation with photometric galaxy surveys from the Rubin Observatory, extending measurements up to a redshift of $\rm{z \sim 2.5}$. This approach can reveal potential deviations from models affecting GW propagation using numerous dark standard sirens in synergy with DESI and the Rubin Observatory.

\end{abstract}


\section{Introduction}
\label{sec:Intro}

The General Theory of Relativity (GR) has been a foundational element in our understanding of gravity for over a century, offering an elegant description of how mass interacts with spacetime curvature. While GR has proven remarkably successful, it encounters challenges under extreme conditions such as those near black holes and during the early universe \citep{thorne1995gravitational, sathyaprakash2009physics}. Recently, a significant breakthrough occurred with the direct detection of gravitational waves (GW) by the LIGO-Virgo collaboration \citep{abbott2016observation, abbott2016gw151226, scientific2017gw170104, abbott2017gw170817, goldstein2017ordinary}. This milestone has opened new pathways for testing GR, especially within the context of compact binary systems in relativistic regimes. These systems, including Binary Neutron Stars (BNS), Neutron Star-Black Hole pairs (NSBH), and Binary Black Holes (BBH), offer unique opportunities to study gravity's fundamental nature across various mass scales and cosmic distances. Ground-based detectors like LIGO \citep{aasi2015advanced}, Virgo \citep{acernese2014advanced}, and KAGRA \citep{akutsu2021overview}, operating in the hertz to kilohertz frequency range, have been crucial in this endeavor. The forthcoming addition of space-based detectors such as LISA \citep{amaro2017laser}, which will function in the millihertz frequency range, alongside advanced ground-based observatories like LIGO-Aundha (LIGO India)\citep{saleem2021science}\footnote{\url{https://dcc.ligo.org/LIGO-M1100296/public}}, Cosmic Explorer (CE) \citep{reitze2019cosmic}, and the Einstein Telescope (ET) \citep{punturo2010einstein}, promises to significantly enhance our capacity to test GR's predictions and deepen our grasp of gravitational phenomena. By leveraging these combined observational capabilities, we aim to refine and expand our understanding of the intricate relationship between gravity and the cosmos. While various alternative models have been proposed to address specific scenarios, this study focuses on testing GR through a model-independent, data-driven approach to GW propagation. Through this rigorous analysis, we seek to explore and validate the fundamental principles of GR across a wide range of conditions and scales. The propagation speed of GW and EM signals, along with their corresponding luminosity distances, can provide valuable insights into modified gravity theories. Deviations from general relativity can arise due to factors such as the mass of the graviton, a frictional term ($\rm{\gamma(z)}$) associated with the running of the Planck mass, and an anisotropic source term \citep{deffayet2007probing, saltas2014anisotropic, nishizawa2018generalized, belgacem2018gravitational, belgacem2018modified, lombriser2016breaking, lombriser2017challenges}.

Various model-dependent parameterizations of $\rm{\gamma(z)}$ have been proposed in the literature. One such parameterization is the $\rm{(\Xi_0, n)}$ model \citep{belgacem2018modified}, which has been shown to be inferable with LVK \citep{Mukherjee:2020mha,leyde2022current, Finke:2021aom, Chen:2023wpj} and LISA \citep{Baker:2020apq}. This parameterization relies on two positive parameters, $\rm{\Xi_0}$ and $\rm{n}$, where setting $\rm{\Xi_0 = 1}$ corresponds to General Relativity (GR). Numerous modified gravity models predict the frictional term under different scenarios. A prominent class among these is Scalar-Tensor theories, which include a scalar degree of freedom crucial for the universe's evolution. Notable scalar-tensor theories include Brans-Dicke theory \citep{brans1961mach}, $f(R)$ gravity \citep{hu2007models}, and covariant Galileon models \citep{chow2009galileon}, all of which fall under Horndeski theories. Beyond Horndeski theories, Degenerate Higher Order Scalar-Tensor (DHOST) theories \citep{frusciante2019tracker, crisostomi2016extended, achour2016degenerate} have been developed as the most general scalar-tensor theories propagating a single scalar degree of freedom along with the helicity-2 mode of a massless graviton.

There are other well-known modified gravity theories in the literature, such as $f(Q)$ gravity, $f(T)$ gravity \citep{cai2016f}, bigravity \citep{de2021minimal}, and gravity in extra dimensions \citep{dvali20004d, yamashita2014mapping, corman2021constraining}. Additionally, there are other parameterizations, such as the polynomial-exponential and exponential parameterizations \citep{belgacem2018modified}, which differ from the power-law form described by $\rm{\Xi_0}$ and $\rm{n}$ \citep{belgacem2019testing}. It is noted that the fitting formula for $\rm{\gamma(z)}$ becomes less accurate at both low and high redshifts compared to some models (see, for example, \citep{deffayet2007probing}). Beyond this limited range of models, the integrated effect of $\rm{\gamma(z)}$ might deviate from a simple power-law form.

To address these challenges, it is essential to test GR in a model-independent manner and measure the integrated effect of $\rm{\gamma(z)}$ as a function of the data, introduced for bright siren in our previous analysis \citep{Afroz:2023ndy,Afroz:2024oui}. We achieve this by using the function $\rm{\mathcal{F}(z)}$, a model-independent parameterization of $\rm{\gamma(z)}$ (more details in Section \ref{sec:Formalism}). This function $\rm{\mathcal{F}(z)}$ is designed to capture any deviation from GR and serves as a metric to assess the difference between the distances $\mathrm{D_l^{GW}(z)}$ and $\mathrm{D_l^{EM}(z)}$. In GR, $\rm{\mathcal{F}(z) = 1}$ at all redshifts, and our approach makes it possible to reconstruct this in a redshift tomographic-way. In this framework, along with $\rm{\mathcal{F}(z)}$, we can also infer the Hubble constant, $\rm{H_0}$ making it possible for jointly explore cosmology and modified gravity theories.

In our previous work, using bright standard sirens (GW sources with EM counterparts), we demonstrated a model-independent reconstruction of the frictional term for various sources, utilizing both current and upcoming ground-based detectors \citep{Afroz:2023ndy,Afroz:2024oui}. However, our previous analysis was limited to BNS and NSBH systems, as these are the most likely to have EM counterparts. Detecting EM counterparts in BNS and NSBH mergers presents significant challenges due to factors such as prompt collapse, ejecta mass, and orientation. While the detectability of collimated jets, such as those associated with GRBs, depends on their opening angles, wider angles increase visibility. It is important to note that kilonovae, which are also produced in these mergers, are isotropic. Thus, inclination isn't directly responsible for the lack of redshift data from kilonovae. However, other factors like distance, instrument sensitivity, and rapid event localization still complicate the detection of EM counterparts, with phenomena such as "structured jets" and "off-axis viewing" further impacting observations \citep{rees1998refreshed, ramirez2001quiescent, kumar2000some}. For the majority of BNS and NSBH events, EM counterparts may not be detectable, especially if baryonic matter is absent in the environments of stellar-mass binary black holes. Notably, the most common class of binary mergers, BBHs, are unlikely to exhibit EM counterparts. Consequently, like other bright siren analyses, our earlier study was confined to sources from which EM counterparts are anticipated, such as BNS and NSBH mergers, provided there is dedicated EM follow-up. To address these limitations, in this study, we have adopted a method that facilitates the exploration of alternative theories of gravity using redshift unknown GW sources. This technique exploits the three-dimensional spatial clustering of GW sources with galaxy redshift surveys to determine their host redshifts through cross-correlation\citep{mukherjee2018beyond,Mukherjee:2019wcg,Mukherjee:2020hyn,Bera:2020jhx,Mukherjee:2020mha,Oguri:2016dgk,Diaz:2021pem,Ghosh:2023ksl}.

\begin{figure}
\centering
\includegraphics[height=4.5cm, width=8.5cm]{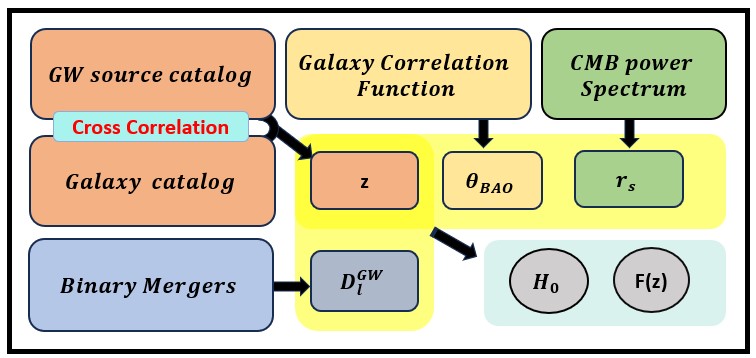}
\caption{This schematic diagram illustrates the measurement of three distinct length scales: the sound horizon ($\rm{r_s}$), the BAO scale ($\rm{\theta_{BAO}}$), and the GW luminosity distance ($\rm{D_l^{GW}}$), along with the redshift of the GW sources. These measurements are obtained from different independent observables, such as CMB observations, galaxy correlation functions, and gravitational wave sources, with the source redshift inferred from the cross-correlation of GW sources and galaxy catalogs. The combination of the sound horizon, the BAO scale, and redshift provides the EM angular diameter distance ($\rm{D^{EM}_A}$). By using the distance duality relation, we can perform a data-driven test of General Relativity as a function of redshift (denoted by $\rm{\mathcal{F}(z)}$) and Hubble constant $\rm{H_0}$.}
\label{fig:motivation}
\vspace{-0.5cm}
\end{figure}

The paper is structured as follows: In Section \ref{sec:Formalism}, we explore the propagation of gravitational waves beyond General Relativity. Section \ref{sec:Mocksamples} shifts focus to the astrophysical population of binary black holes. In Section \ref{sec:RedshiftInf}, we present a detailed framework for inferring redshift using cross-correlation techniques with galaxies of known redshifts. Section \ref{sec:BAO} offers an in-depth discussion on the model-independent measurement of the BAO scale from the galaxy power spectrum. Section \ref{sec:Forecast} elaborates on the formalism underpinning our work and its results. Finally, Section \ref{sec:conclusion} summarize our key findings and discuss future prospects.

\section{Formalism: Non-parametric reconstruction of deviation from General Relativity}
\label{sec:Formalism}

The propagation of GW in the fabric of spacetime, as described by the GR, can be mathematically expressed as follows
\begin{equation}
\mathrm{h_{(+,\times)}^{''} + 2\mathcal{H}h_{(+,\times)}^{'} + c^2k^2h_{(+,\times)} = 0},
\end{equation}
where $\rm{h_{(+,\times)}}$ is GW strain of plus ($\rm{+}$) and cross ($\rm{\times}$) polarization, the prime symbol denotes the derivative with respect to the conformal time($\rm{\eta}$) and $\rm{\mathcal{H}}$ represents the Hubble parameter in comoving coordinates. We utilize this equation as a starting point to examine the propagation of GW and evaluate the validity of the GR. The modified theory of gravity generalizes this equation as follows \citep{Nishizawa:2017nef}
\begin{equation}
\mathrm{h_{(+,\times)}^{''} + 2(1-\gamma(z))\mathcal{H}h_{(+,\times)}^{'} + (c_{GW}^2k^2+m_{GW}^2a^2)h_{(+,\times)} = a^2\Pi_{(+,\times)}}. 
\label{eq:main}
\end{equation}
In this formulation, several additional parameters come into play. Firstly, we have the frictional term denoted by $\rm{\gamma(z)}$, which represents the influence of friction in the theory. The speed of GW propagation is represented by $\rm{c_{GW}}$, a is the scale factor and the graviton mass by  $\rm{m_{GW}}$. Finally, the anisotropic stress term is denoted by $\rm{\Pi_{(+,\times)}}$. It is noteworthy that, within the framework of the GR, all these additional parameters assume a fiducial value of zero, except for $\rm{c_{GW}}$, which corresponds to the speed of light($\rm{c}$). In the context of testing GR through propagation, it is essential to scrutinize the additional parameters and ascertain whether they deviate from the values predicted by GR. Recent measurements, particularly from the GW170817 event, have provided stringent constraints, indicating that the graviton mass ($\rm{m_{GW}}$) is zero, and the speed of GW propagation ($\rm{c_{GW}}$) is equivalent to the speed of light ($\rm{c}$) \citep{abbott2017gravitational, abbott2019tests}. A modification of the the anisotropic stress term ($\rm{\Pi_{+,\times}}$) affects the phase of the binary waveforms; the recent BBH observations, in particular of GW150914 and GW151226, have set some limit on such modifications, although for the moment not very stringent \citep{abbott2016binary}.

The presence of the frictional term $\rm{\gamma(z)}$ changes the amplitude of the GW signal received from a source at cosmological distance. This is particularly interesting because it implies that the luminosity distance measured with standard sirens is in principle different from that measured with standard candles or other electromagnetic (EM) probes such as Cosmic Microwave Background (CMB) or Baryon Acoustic Oscillation (BAO), and this could provide a “smoking gun” signature of modified gravity. So as a consequence the GW luminosity distance $\rm{D_l^{GW}}$ is related to the EM wave-based luminosity distance $\rm{D_l^{EM}}$ by an exponential factor, where the exponent involves integrating $\rm{\gamma(z')/(1+z')}$ with respect to $z'$ \citep{belgacem2018gravitational}. Consequently, the expression becomes
\begin{equation}
\mathrm{D_{l}^{GW}(z) = \exp\left(-\int dz'\frac{\gamma(z')}{1+z'}\right) D_l^{EM}(z)}.
\label{eq:gamma_z}
\end{equation}

In our investigation, we focus on tensor perturbations, where the impact is encoded in the non-trivial function $\rm{D_{l}^{GW}(z)/D_l^{EM}(z)}$. Following our previous works \citep{Afroz:2023ndy, Afroz:2024oui} a non-parametric reconstruction of the deviation from GR as
\begin{equation}
\mathrm{D_{l}^{GW}(z)={\mathcal{F}}(z)D_l^{EM}(z)}.
\label{eq:f(z)}
\end{equation}
The term $\rm{\mathcal{F}(z)}$ can capture any deviation from GR as a function of redshift. The reconstruction of this quantity from observation of $\rm{D_{l}}^{GW}(z)$ and $\rm{D_{l}}^{EM}(z)$ can capture any modified gravity models which predict a modification in the GW propagation \citep{belgacem2019testing}. Alternatively, a parametric form using $\rm{\Xi_0}$ and $\rm{n}$, $\rm{D_{l}^{GW}(z)/D_l^{EM}(z)=\Xi_0+\frac{1-\Xi_0}{(1+z)^n}}$ is also used  to capture all those models which predicts a power-law modification with redshift of the GWs propagation effect \citep{belgacem2018modified}. However, in this parametric form the value of $n$ cannot be constrained for GR ($\rm{\Xi_0=1}$), as the second term vanishes. The model-independent approach also makes it possible to avoid this issue. 

Several alternative models of GR predict deviation in different natures of the acceleration of the Universe that are predicted from the $\Lambda$CDM. As a result, this data-driven approach can make it possible to explore both deviation from GR and $\rm{w=-1}$ equation of state of dark energy. The use of different observational probes was done previously \citep{mukherjee2021testing} for a parametric form of deviation from GR (using $\rm{\Xi_0}$ and $\rm{n}$) valid for a class of model. However, the non-parametric form using $\rm{\mathcal{F}(z)}$ proposed in this analysis, can make a redshift-dependent reconstruction of any deviation from GR using the multi-messenger observations. In addition to these methods, the bright siren approach has been employed to demonstrate model-independent deviations from GR. This has been achieved using ongoing and upcoming ground-based detectors \citep{Afroz:2023ndy} as well as upcoming space-based detectors such as LISA \citep{Afroz:2024oui}. These studies highlight the potential of bright sirens to provide independent constraints on deviations from GR, complementing other observational probes.

\subsection{Data-Driven Reconstruction of F(z)}

\textbf{Inference of GW luminosity distance:} In the realm of binary systems, GW offer a unique method for determining the luminosity distance, denoted as $\rm{D_{l}^{GW}}$. This distance measurement is derived from the analysis of GW strain emitted by events during the inspiral and merger of compact objects like NSBH, BNS, and BBH systems. The amplitude of the GW signal, which diminishes with distance, provides a direct measure of $\rm{D_{l}^{GW}}$. By analyzing the waveform of these GW events, which contains detailed information about the binary system's parameters such as chirp mass and orbital inclination, we can accurately extract the luminosity distance.

\textbf{Inference of EM luminosity distance:} A data-driven approach to infer the EM luminosity distance can be made from the angular diameter distance $\rm{D^{EM}_A(z)}$ measured using BAO angular peak position ($\rm{\theta_{BAO}(z)}$) from large scale structure observation  at different redshifts \citep{peebles1973statistical}, and using the distance duality relation which connects the angular diameter distance with the EM luminosity distance by $\rm{D^{EM}_A(z)= D^{EM}_l(z)/(1+z)^2}$. The position of the BAO peak in the galaxy two-point correlation function can be inferred directly from observation. The angular scales are related to the comoving sound horizon until the redshift of drag epoch ($\rm{z_d \sim 1020}$) as 

\begin{equation}
\mathrm{r_s=\int_{z_d}^{\infty}\frac{dz\,c_s(z)}{H_0\sqrt{\Omega_m(1+z)^3+\Omega_\Lambda + \Omega_r(1+z)^4}}}.
\end{equation}

By replacing $\rm{D_l^{EM}}$ in Equation \eqref{eq:f(z)} with the $\rm{\theta_{BAO}(z)}$ and by using the distance duality relation for EM probes, one can write the above equation as \citep{mukherjee2021testing}
\begin{equation}\label{eq:f}
\mathrm{\mathcal{F}(z)=\frac{D_l^{GW}(z)\theta_{BAO}(z)}{(1+z)r_s}}.
\end{equation}
The comoving sound horizon $\rm{r_s}$ at the drag epoch can be related to the comoving sound horizon $\rm{r_*}$ at the redshift of recombination ($\rm{z_*= 1090}$) by $\rm{r_d \approx 1.02r_*}$. The quantity $r_*$ is inferred from CMB observation, using the position of the first peak in the CMB temperature power spectrum $\rm{\theta_*= r_*/(1+z_*)D^{EM}_A(z_*)}$ \citep{spergel2007three, hinshaw2013nine, aghanim2020planck}. As a result, the product of all these observable quantities will lead to an identity value at all redshifts if GR and the fiducial $\Lambda$CDM model of cosmology is the correct theory. However, if there is any departure from these models, then we can measure a deviation from one with redshift. It is important to note that the value of $\rm{r_s}$ is constant in the redshift range relevant for this study ( $\rm{z<2}$). So, even if there is an inaccuracy in the inference of true $\rm{r_s}$, the value of $\rm{\mathcal{F}(z)}$ will vary by an overall normalization. But the reconstructed redshift evolution of $\rm{\mathcal{F}(z)}$ will remain the same. In one of the later sections, we will show results for both the cases, (i) only $\rm{\mathcal{F}(z)}$ and (ii) $\rm{\mathcal{F}(z)}$ and $\rm{H_0}$ together. The second case will make it possible to marginalise over any inaccurate inference of sound horizon $\rm{r_s}$  due to incorrect inference of the Hubble constant. In the above equation, $\rm{D_l^{GW}(z)}$ is measured from GW sources, $\rm{\theta_{BAO}}$ is measured from galaxy two point correlation function, and $\rm{r_s}$ is inferred from the CMB temperature fluctuations. The redshift of the source can be inferred using cross-correlation of galaxy surveys with dark sirens. In Subsection \ref{sec:RedshiftInf}, we provide a detailed framework for inferring redshift using cross-correlation with redshift-known galaxies.

\section{GW mock samples for LVK, Cosmic Explorer, and Einstein Telescope}
\label{sec:Mocksamples}
The study of GW has been significantly advanced by the efforts of the LIGO-Virgo-KAGRA (LVK) collaboration, which consists of ground-based detectors designed to observe cosmic events such as binary mergers. The LVK detectors include the Advanced LIGO detectors in the United States and the Virgo detector in Italy, along with the KAGRA detector in Japan. These detectors operate at frequencies between 10 Hz to a few kHz, and have already provided critical insights into the nature of compact objects and the expansion of the universe.

Looking ahead, next-generation ground based detectors such as the Cosmic Explorer (CE) \citep{Evans:2021gyd} and the Einstein Telescope (ET) \citep{Branchesi:2023mws} are set to further revolutionize the field. The CE is planned for construction in the United States, will feature a 20-kilometer(or/and 40-kilometer) arm length, significantly improving sensitivity and the ability to detect events at greater distances and lower frequencies compared to current detectors. The ET is a European project, will have a unique triangular configuration with 10-kilometer-long arms and will be built underground to reduce seismic noise. The ET will cover a frequency range from a few Hz to several kHz, enabling it to detect a wider variety of astrophysical sources. Together, these advanced detectors will enhance our understanding of GW and provide new opportunities to test the predictions of GR and alternative theories of gravity.

\subsection{Modelling of the astrophysical population of GW source}
To accurately assess deviations from the GR model, it is crucial to rely on real GW events. The total number of GW events is contingent upon the merger rates, and mass population of the GW sources. We will explain below the astrophysical models used in the analysis. 

\texttt{Merger rate: } In this study we use a delay time model of the binary mergers  \citep{o2010binary, dominik2015double}. In the delay time model, the merger rate is described in terms of the delay time distribution, denoted as $\rm{t_d}$. The delay time refers to the elapsed time between the formation of stars that will eventually become black holes and the actual merging of these black holes. It is important to note that the time delay is not uniform across all binary black holes but instead follows a specific distribution. This distribution function accounts for the variations in the delay time and is defined as follows
\begin{equation}
    \mathrm{p_t(t_d|t_d^{min},t_d^{max},d) \propto 
    \begin{cases}
    (t_d)^{-d} & \text{, for }  t_d^{min}<t_d<t_d^{max}, \\
    0 & \text{otherwise}.
    \end{cases}}
\end{equation}
The delay time is given by $\rm{t_d=t_m-t_f}$, where $\rm{t_m}$ and $\rm{t_f}$ are the lookback times of merger and formation respectively \citep{karathanasis2023binary}. So the merger rate at redshift $\rm{z}$ can be defined as
\begin{equation}
    \mathrm{R_{TD}(z)=R_0\frac{\int_z^{\infty}p_t(t_d|t_d^{min},t_d^{max},d)R_{SFR}(z_f)\frac{dt}{dz_f}dz_f}{\int_0^{\infty}p_t(t_d|t_d^{min},t_d^{max},d)R_{SFR}(z_f)\frac{dt}{dz_f}dz_f}}. 
\end{equation}
The parameter $\rm{R_0}$ represents the local merger rate, indicating the frequency of mergers at a redshift of $\rm{z=0}$. According to the study in \citep{abbott2023population}, the estimated values of $\rm{R_0}$ for BBH merger rate to be between 17.9 $\rm{Gpc^{-3}yr^{-1}}$ and 44 $\rm{Gpc^{-3}yr^{-1}}$ at a fiducial redshift ($\rm{z = 0.2}$). In our study, we assume a standard local merger rate of $\rm{R_0}$ = 20 $\rm{Gpc^{-3} yr^{-1}}$ for the BBH system. The numerator of the expression involves the integration over redshift $\rm{z_f}$ from $\rm{z}$ to infinity, where $\rm{p_t(t_d|t_d^{min},t_d^{max},d)}$ is the delay time distribution, $\rm{R_{SFR}(z_f)}$ is the star formation rate, and $\rm{\frac{dt}{dz_f}}$ is the jacobian of the transformation. The star formation rate($\rm{R_{SFR}(z)}$), is determined using the \citep{madau2014cosmic} star formation rate.

The total number of compact binary coalescing events per unit redshift is estimated as 
\begin{equation}
\mathrm{\frac{dN_{GW}}{dz} = \frac{R_{\rm TD}(z)}{1+z} \frac{dV_c}{dz} T_{obs}},
\label{eq:totno}
\end{equation}
where $T_{obs}$ indicates the total observation time, $\rm{\frac{dV_c}{dz}}$ corresponds to the comoving volume element, and $\rm{R(z)}$ denotes the merger rate \citep{karathanasis2022gwsim}. We consider the delay time merger rate with a specific minimum delay time $\rm{t_d =500\text{Myrs}}$ and a power-law exponent of $\rm{d=1}$\citep{2010MNRAS.402..371B,Dominik:2014yma,Cao:2017ztl}. To determine which events are detectable, the calculation of the matched filtering signal-to-noise ratio (SNR) plays a crucial role. The SNR serves as a measure of the strength of the GW signal relative to the background noise. Only those events with a matched filtering SNR greater than or equal to a predetermined threshold SNR ($\rm{\rho_{\rm TH}}$) can be reliably detected. \citep{maggiore2007gravitational}.  

For a GW emitted by an optimally oriented binary system, the optimized SNR, denoted as $\rm{\rho}$, is defined as follows \citep{sathyaprakash1991choice, cutler1994gravitational, balasubramanian1996gravitational, nissanke2010exploring}

\begin{equation}
\mathrm{\rho^2 \equiv 4\int_{f_{\text{min}}}^{f_{\text{max}}} df \frac{|h(f)|^2}{S_n(f)}},
\label{snr}
\end{equation}
here, $\rm{S_n(f)}$ represents the power spectral density of the detector. The function $\rm{h(f)}$ corresponds to the GW strain in the restricted post-Newtonian approximation and is defined for plus ($\rm{+}$) and cross ($\rm{\times}$) polarization as \citep{ajith2008template} 

\begin{equation}
    \mathrm{h(f)_{\{+, \times\}}=\sqrt{\frac{5\eta}{24}}\frac{(GM_c)^{5/6}}{D_l\pi^{2/3}c^{3/2}}f^{-7/6}e^{\iota\Psi(f)}\mathcal{I}_{\{+, \times\}}}.
\end{equation}

In this expression, the symbol $\rm{\eta}$ represents the symmetric mass ratio. The term $\rm{M_c}$ signifies the chirp mass of the system. The variable $\rm{D_l}$ denotes the luminosity distance. The constant c represents the speed of light in a vacuum. $\rm{\mathcal{I}_{+}= (1+\cos^{2}i)/2}$ and $\rm{\mathcal{I}_{\times}= \cos i}$ depends on the inclination angle $\rm{i}$. Finally, $\rm{\Psi(f)}$ stands for the phase of the waveform. However, the signal detected by a GW detector $\rm{h_{det}}$ is a complex interplay of several variables, including the detection antenna functions ($\rm{F_{+}, F_{\times}}$), and can be expressed as 
\begin{equation}
    \mathrm{h_{det}=F_{+}h_{+}+F_{\times}h_{\times}},
\end{equation}
here $\rm{F_{+}}$ and $\rm{F_{\times}}$ are the antenna functions defined as follows \citep{finn1993observing}
\begin{align}\label{eq:antenna}
    &\mathrm{F_{+}=\frac{1}{2}(1+cos^2\theta)cos2\phi cos2\psi-cos\theta sin2\phi sin2\psi},\\\nonumber
     &\mathrm{F_{\times}=\frac{1}{2}(1+\cos^2\theta)\cos2\phi sin2\psi+cos\theta sin2\phi cos2\psi},
\end{align}
where $\rm{\theta}$ and $\rm{\phi}$ define the location of the source in the sky, and $\rm{\psi}$ is related to the orientation of the binary system with respect to the detector. Consequently, the matched filtering SNR ($\rho$) takes the form \citep{finn1996binary}

\begin{equation}
    \mathrm{\rho = \frac{\Theta}{4}\biggl[4\int_{f_{min}}^{f_{max}}h(f)^2/S_n(f)df\biggr]^{1/2}},
    \label{eq:SNR}
\end{equation}
where $\rm{\Theta^2 \equiv 4 \left(F_{+}^2(1+\cos^2i)^2 + 4F_{\times}^2\cos^2i\right)}$. Averaging over many binaries inclination angle and sky positions, \citep{finn1996binary} showed that $\Theta$ follows a distribution

\begin{equation}
\mathrm{P_{\Theta}(\Theta) = 
\begin{cases}
5\Theta(4-\Theta)^3/256& \text{if } 0<\Theta<4,\\
0,              & \text{otherwise}.
\end{cases}}
\end{equation}

\texttt{Mass model:} The mass population for BBH is informed by the recent findings from the third catalog of GW sources published by the LVK collaboration \citep{abbott2023population, talbot2018measuring, abbott2019binary}. To accurately model the mass distribution of black holes, we utilize the Power Law + Gaussian Peak model. This approach integrates a power law distribution for the initial mass distribution and adds a Gaussian component to capture the concentrated occurrences of black hole masses ranging between $\rm{5M_{\odot}}$ and $\rm{50M_{\odot}}$, with additional smoothing to ensure a realistic distribution.

In this study, we employ two sets of GW detectors: one comprising LIGO \citep{aasi2015advanced}, Virgo \citep{acernese2014advanced}, and KAGRA \citep{akutsu2021overview} (LVK), and the other consisting of CE with two different configurations: a 20 km configuration (CE(20Km)+ET) and a 40 km configuration (CE(40Km)+ET), along with ET \citep{punturo2010einstein}. We establish different SNR thresholds for different GW observatories: $\rm{\rho_{\rm Th} = 12}$ for the LVK (O5 sensitivity) system and $\rm{\rho_{\rm Th} = 20}$ for both configurations of the CE+ET. We adopt a redshift bin size of $\rm{\Delta z = 0.05}$, within which the total number of events is computed using Equation \eqref{eq:totno}. Event generation is based on distances derived from redshifts, parameter $\rm{\Theta}$, and the mass distributions of the objects. Over a five-year operational period with a 75\% duty cycle, the expected number of detectable events is projected to be approximately $\sim$13000 for LVK, $\sim$177000 for CE(40Km)+ET, and $\sim$144000 for CE(20Km)+ET, assuming a local merger rate ($\rm{R_0}$) of 20 $\rm{Gpc^{-3} yr^{-1}}$. It is important to note that for this analysis, we have specifically utilized the CE(20 km)+ET configuration. From now onwards, this combination is defined as CEET.

\begin{figure}
\centering
\includegraphics[height=5.5cm, width=9cm]
{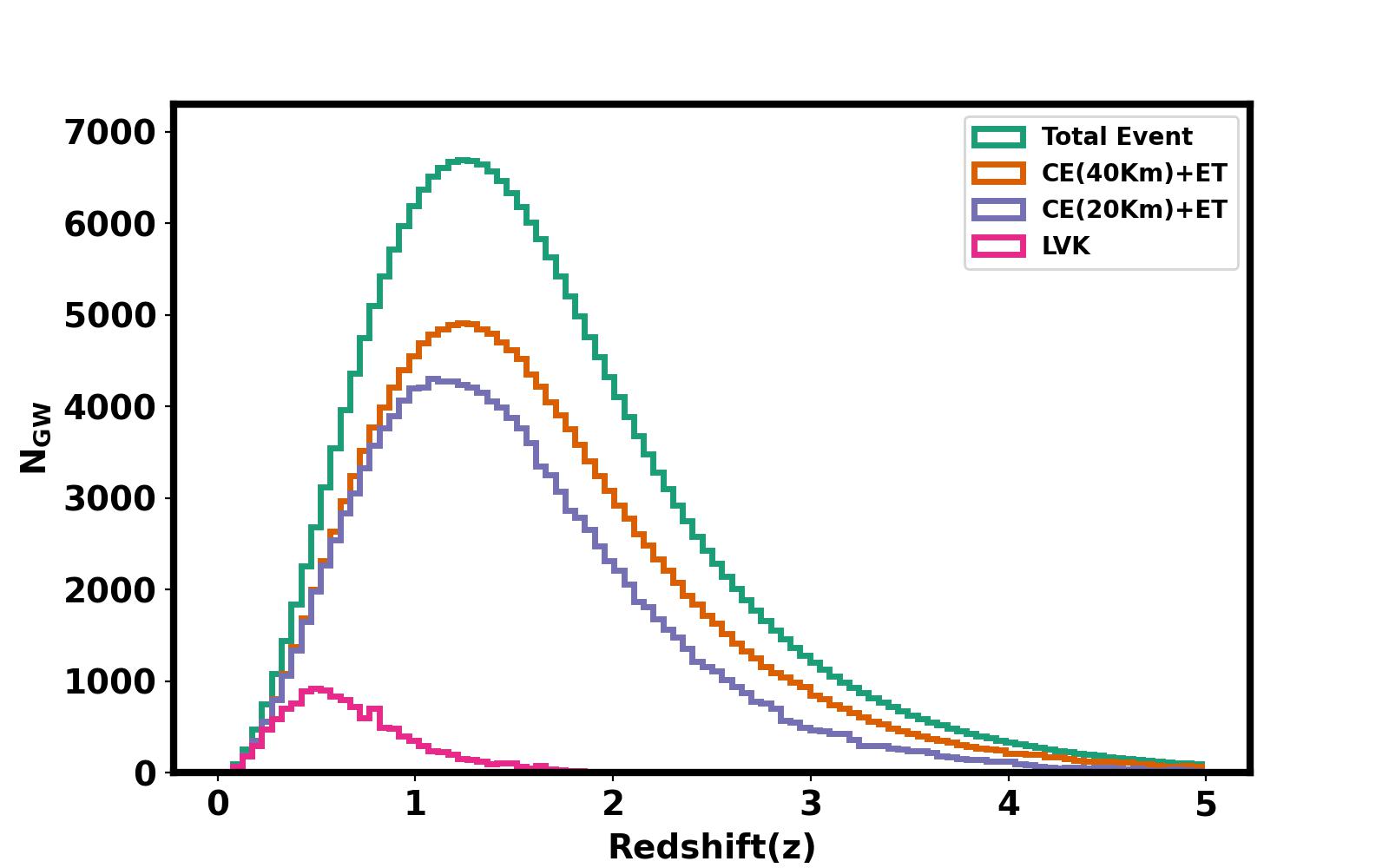}
\caption{The figure depicts the total number of injected events (in black) and the number of detectable GW events as a function of redshift for binary black hole (BBH) sources. The detections are shown for the CE(40Km)+ET (in blue), CE(20Km)+ET (in green), and the LVK system (in golden). Each observation period is assumed to be 5 years with a 75\% duty cycle.}
\label{fig:GWevent}
\end{figure}

In Figure \ref{fig:GWevent}, we present the projected total and detectable BBH events within specific redshift intervals ($\rm{\Delta z = 0.05}$). To obtain these estimates, we first calculate the number of BBH mergers for each redshift bin using Equation \eqref{eq:totno}. For each BBH event, we use the inverse transform method to determine the mass of the two black holes and the parameter $\rm{\Theta}$, drawing from their respective probability distributions. The masses of the black holes are generated within the range of $\rm{5M_{\odot}}$ to $\rm{50M_{\odot}}$, and $\rm{\Theta}$ is considered within the interval from 0 to 4. After incorporating the redshift information for each event, we calculate the corresponding luminosity distance and the SNR using Equation \eqref{eq:SNR} for a single detector. The total SNR, denoted as $\rm{\rho_{total}}$, is calculated by combining the individual SNRs from the network of detectors using the formula $\rm{\rho_{total}=\sqrt{\sum_i \rho_i^2}}$. This analysis considers the detection capabilities of both the CEET and the LVK network over a specified observation period of 5 years with a 75\% duty cycle. The curve depicted in the graph illustrates the distribution of these events, providing insights into the temporal occurrence of these mergers through cosmic history and their detection likelihood with current and future GW observatories.

\subsection{GW Source Parameter Estimation using Bilby}
\label{sec:ParamEstBilby}

We initiate the parameter estimation process for the identified sources using the \texttt{Bilby} package \citep{ashton2019bilby}, which provides realistic posterior distributions of the GW luminosity distance, marginalized over other source parameters. The masses of the GW sources and their number at different redshifts are determined according to methods described previously.

\begin{figure}
\centering
\includegraphics[height=5.5cm, width=9cm]{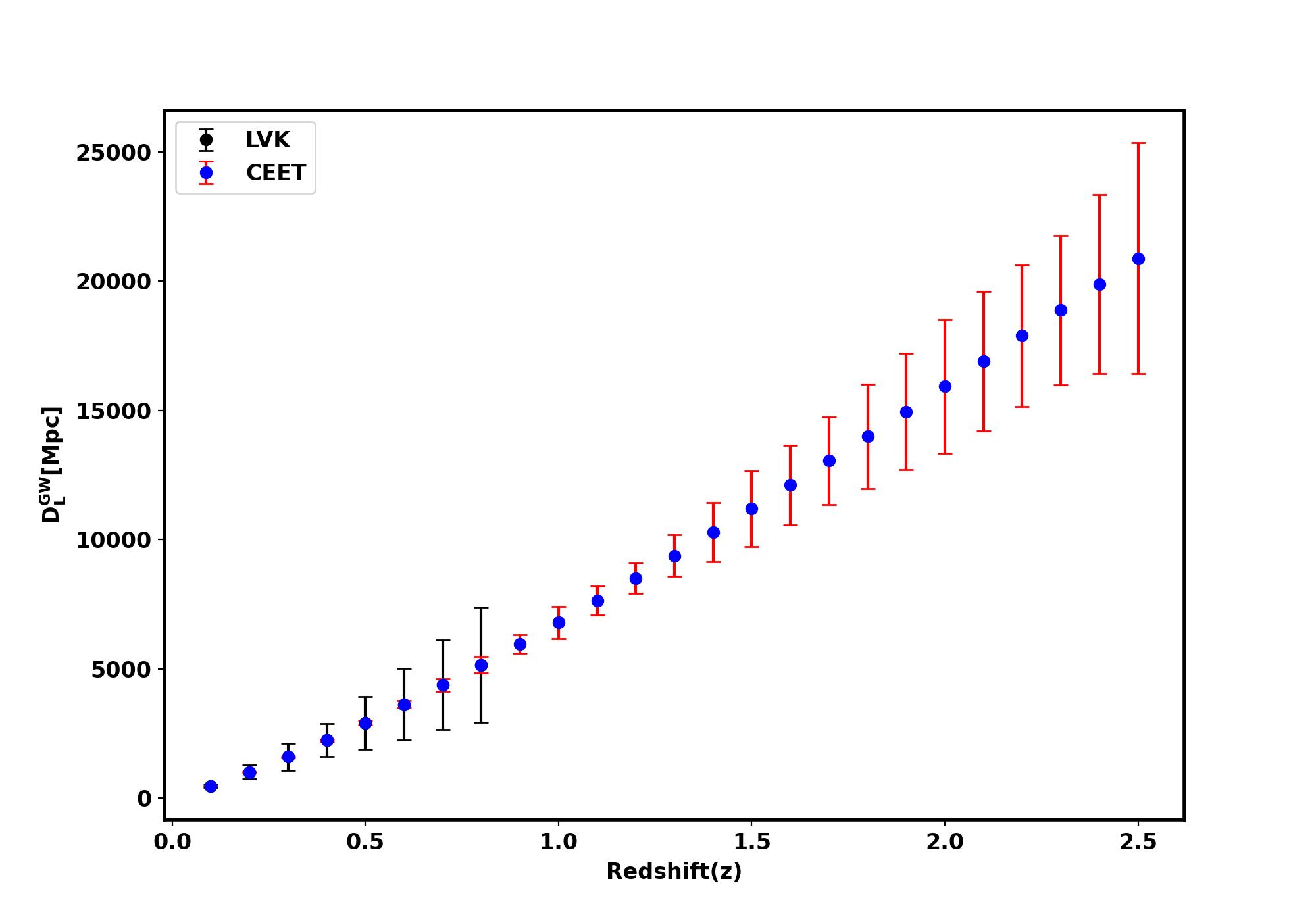}
\caption{This figure compares the luminosity distance errors as a function of redshift for both the CEET and LVK networks. Each data point represents the measured luminosity distance along with the uncertainty derived from GW source parameter estimation. It is evident that future detectors, such as CEET, are expected to measure these distances with greater precision.}
\label{fig:dist}
\end{figure}

For additional source parameters such as the inclination angle ($\rm{i}$), polarization angle ($\rm{\psi}$), GW phase ($\rm{\phi}$), right ascension (RA), and declination (Dec), we adopt uniform sampling, assuming a non-spinning system. We then generate a GW signal using the \texttt{IMRPhenomHM} waveform model \citep{kalaghatgi2020parameter}, which includes higher-order modes to help reduce the degeneracy between the luminosity distance ($\rm{D_l^{GW}}$) and the inclination angle ($\rm{i}$). By fixing the priors for all parameters except $\rm{m_1}$, $\rm{m_2}$, $\rm{D_l^{GW}}$, $\rm{i}$, RA, and Dec to delta functions, we obtain detailed posterior distributions. Among these, the posterior distribution for $\rm{D_l^{GW}}$ is particularly vital for our analysis as it informs the inference of $\rm{\mathcal{F}(z)}$, while RA and Dec are essential for estimating sky localization errors. In Figure \ref{fig:dist}, we present a comparison of luminosity distance errors as a function of redshift for two GW detector networks: CEET and LVK. Each data point represents the measured luminosity distance along with the uncertainty derived from GW source parameter estimation. The plot demonstrates that, overall, CEET is expected to achieve better precision in measuring luminosity distances compared to the LVK network. This improvement highlights the advancements in sensitivity and accuracy expected from future GW detectors.

\section{Cross-Correlation Technique for Redshift Estimation of Dark Sirens}
\label{sec:RedshiftInf}

The large-scale matter distribution in the Universe is statistically homogeneous and isotropic, in accordance with the Copernican principle. This enables us to describe the large-scale distribution of galaxies using the galaxy density field $\rm{\delta_g(r)}$, defined as
\begin{equation}
    \mathrm{\delta_g(r) = \frac{n_g(r)}{\bar{n}_g} - 1},
\end{equation}
where \(\rm{n_g(r)}\) is the number density of galaxies at position \(\rm{r}\), and \(\rm{\bar{n}_g}\) is the mean number density of galaxies. In the standard cosmological model, the spatial distribution of galaxies traces the underlying matter distribution in the Universe and can be represented as a biased tracer of the matter density field \(\rm{\delta_m(k,z)}\) with the relation 
\begin{equation}
    \mathrm{\delta_g(k,z) = b_g(k,z) \delta_m(k,z)},
\end{equation}
here \(\rm{b_g(k,z)}\) is the galaxy bias parameter, and \(\rm{\delta_g(k,z)}\) is the Fourier transform of the real-space galaxy density field \(\rm{\delta_g(r,z)}\). The galaxy bias parameter \(\rm{b_g(k,z)}\) indicates how galaxies trace the dark matter distribution. Galaxy catalogs from ongoing and upcoming surveys such as DES\citep{DES:2017myr}, DESI\citep{DESI:2016fyo}, Euclid\citep{2010arXiv1001.0061R}, LSST\citep{2009arXiv0912.0201L}, and SPHEREx\citep{SPHEREx:2018xfm} will be available up to redshift $\rm{z=3}$. These surveys combined will cover nearly the full sky, which will enhance the overlap with the GW sources. By providing extensive and detailed data on the distribution of galaxies, these surveys will enhance our understanding of the Universe, allowing us to test cosmological models with greater precision and gain deeper insights into the large-scale structure of the cosmos.

Astrophysical GW events are expected to occur in galaxies, following their spatial distribution. This distribution is characterized by a bias parameter, \(\rm{b_{GW}(k, z)}\), which differs from the galaxy bias parameter, \(\rm{b_g(k, z)}\). The density field for GW sources in real space, \(\rm{\delta_{GW}(k, z)}\), is defined as
\begin{equation}
    \mathrm{\delta_{GW}(k, z) = b_{GW}(k, z) \delta_m(k, z)},
\end{equation}
here, \(\rm{b_{GW}(k, z)}\) describes how GW sources trace the large-scale structure of the Universe. However, GW sources are characterized by their luminosity distance (\(\rm{D_l^{GW}}\)) and sky localization \((\rm{\theta_{GW}, \phi_{GW})}\), introducing a sky localization error \(\rm{\Delta \Omega_{GW}}\). Sky localization error refers to the uncertainty in determining the precise position of a GW source in the sky. This error arises due to the inherent uncertainties in measuring the declination ($\rm{\theta_{GW}}$) and right ascension ($\rm{\phi_{GW}}$)  of the GW source, and it is defined as \citep{Singer:2015ema}

\begin{equation}
    \mathrm{\Delta\Omega_{GW} = \sin(\theta_{GW}) \sqrt{\sigma_{\theta_{GW}}^2 \sigma_{\phi_{GW}}^2 - \sigma_{\theta_{GW} \phi_{GW}}^2}}, 
\end{equation}

here $\rm{\sigma_{\theta_{GW}}}$ represents the error in $\rm{\theta_{GW}}$ (in radians), $\rm{\sigma_{\phi_{GW}}}$ is the error in $\rm{\phi_{GW}}$ (in radians), and $\rm{\sigma_{\theta_{GW} \phi_{GW}}}$ is the covariance between $\rm{\theta_{GW}}$ and $\rm{\phi_{GW}}$. The covariance term $\rm{\sigma_{\theta_{GW} \phi_{GW}}}$ is calculated from the joint distribution of samples of $\rm{\theta_{GW}}$ and $\rm{\phi_{GW}}$ using the formula
\begin{equation}
    \mathrm{\sigma_{\theta_{GW} \phi_{GW}} = \frac{1}{N - 1} \sum_{i=1}^{N} (\theta_{GW_i} - \bar{\theta}_{GW})(\phi_{GW_i} - \bar{\phi}_{GW})}, 
\end{equation}
here $\rm{N}$ is the number of samples, $\rm{\theta_{GW_i}}$ and $\rm{\phi_{GW_i}}$ are individual samples of $\rm{\theta_{GW}}$ and $\rm{\phi_{GW}}$, and $\rm{\bar{\theta}_{GW}}$ and $\rm{\bar{\phi}_{GW}}$ are the means of the sample sets for $\rm{\theta_{GW}}$ and $\rm{\phi_{GW}}$, respectively. This error leads to uncertainty in the precise position of the GW source within the sky localization region, blurring the spatial information. Consequently, the density field of GW sources is modified due to this sky localization error, expressed as
\begin{equation}
    \mathrm{\delta_{\text{GW}}^r(k, \Delta \Omega_{GW}, z) = \delta_{\text{GW}}^r(k, z) e^{-\frac{k^2}{k_{\text{eff}}^2(z)}}},
\end{equation}
where \(\rm{\delta^r_{\text{GW}}(k, z)}\) is the original density field of GW sources without sky localization impact, and \(\rm{k_{\text{eff}}(z)}\) denotes a characteristic wavenumber varying with redshift \(\rm{z}\) \citep{Mukherjee:2020hyn,Mukherjee:2020mha}. \(\rm{k_{\text{eff}}(z)}\) is the comoving scale defined as 
\begin{equation}
    \mathrm{k_{\text{eff}}(z) = \sqrt{\frac{8 \ln 2}{\Delta \Omega_{GW} D_c(z)^2}}},
\end{equation}
where \(\rm{D_c(z)}\) is the comoving distance at redshift $\rm{z}$ \citep{Mukherjee:2020hyn}. The exponential term \(\rm{e^{-\frac{k^2}{k_{\text{eff}}^2(z)}}}\) quantifies the smoothing effect caused by the sky localization error, with a rapid decrease beyond \(\rm{k > k_{\text{eff}}}\). Ongoing GW detectors like LVK are helping us characterize detected GW sources. In the future, upcoming ground-based detectors such as CE, ET, and LIGO-India, along with space-based detectors like LISA, will enhance our ability to characterize these sources. These advancements will improve our measurements of the redshifts of GW sources, thereby providing better insights into the Universe's large scale structure and evolution. 

In the standard model of cosmology, galaxies trace the underlying distribution of dark matter through the galaxy bias parameter, $\rm{b_g(k, z)}$. Consequently, GW sources are also expected to trace the dark matter distribution, but with their own distinct bias parameter, $\rm{b_{GW}(k, z)}$, which differs from the galaxy bias parameter.

The spatial clustering of galaxies and GW sources can also be described in Fourier space using the three-dimensional auto-power spectrum and cross-power spectrum at different redshifts \(\rm{z}\). The matter distribution exhibits a clustering property that can be statistically described by the correlation function \(\rm{\xi(r)}\), which is the fourier transform of the matter power spectrum \(\rm{P_m(k) = \langle \delta_m(k) \delta_m(k) \rangle}\). Under the standard cosmological model, both GW events and the distribution of galaxies are biased tracers of the underlying matter distribution. Therefore, the cross-correlation of the power spectrum between GW sources and galaxy surveys can be used to infer the host redshift shell of the GW sources. For a detailed mathematical formulation, refer to Subsection \ref{sec:MathFormRedInf}.

\subsection{Mathematical Formulation of Cross-Correlation Technique}
\label{sec:MathFormRedInf}

To estimate the redshift ($\rm{z}$) of a GW binary source when its EM is not detectable, we utilize the cross-correlation between galaxy surveys and dark sirens. The relevant probability distribution for this estimation is given by:

\begin{equation}
   \mathrm{\mathcal{P}(z|\vec{\vartheta}_{GW}, \vec d_{g})\propto \mathcal{L}(\vec{\vartheta}_{GW}| P^{ss}_{gg}(\vec k,z), \vec d_g(z)) \mathcal{P}(\vec d_g| P^{ss}_{gg}(\vec k,z))},
\end{equation}

In this expression, $\rm{\vec{\vartheta}_{\text{GW}} = (D_l^{\text{GW}}, \theta_{\text{GW}}, \phi_{\text{GW}})}$ represents the GW data vector, which includes the luminosity distance $\rm{D_l^{GW}}$ to the GW source, as well as the sky localization of the source $(\theta_{\text{GW}}, \phi_{\text{GW}})$. The galaxy data vector $\rm{\vec{d}_g = (\delta_g(z_g, \theta_g, \phi_g))}$ includes the redshift information of the galaxy $\rm{z_g}$ as well as the sky position $(\rm{\theta_g, \phi_g})$. The term  $\rm{\mathcal{L}(\vec{\vartheta}_{\text{GW}} | P^{ss}_{gg}(\vec{k}, z), \vec{d}_g(z))}$ represents the likelihood function, which measures how well the GW data fits the galaxy density field based on a model for the galaxy power spectrum  $\rm{P^{ss}_{gg}(\vec{k}, z)}$. This power spectrum describes the distribution of galaxy densities in redshift space, accounting for both cosmological redshift and additional redshift effects due to the peculiar velocities of galaxies. The term $\rm{\mathcal{P}(\vec{d}_g | P{gg}^{ss}(\vec{k}, z))}$ denotes the posterior distribution of the galaxy density field given the power spectrum $\rm{P^{ss}_{gg}(\vec{k}, z)}$. It provides the probability of observing the galaxy data under the assumed model for the galaxy power spectrum. The superscript 's' indicates that the galaxy survey in question is either a spectroscopic or photometric survey, observing galaxies in redshift space, which includes both cosmological redshift and redshift due to peculiar velocities. Overall, the probability distribution,  $\rm{\mathcal{P}(z | \vec{\vartheta}_{GW}, \vec{d}_g)}$  combines information from GW data and galaxy surveys to estimate the redshift of the GW source. The posterior distribution of the galaxy density field, conditional on the galaxy power spectrum, is described by the equation

\begin{equation}
\mathrm{\mathcal{P}(\vec d_g| P^{ss}_{gg}(\vec k,z)) \propto \exp\bigg(-{\frac{ \delta^{s}_g(\vec k, z) \delta^{s*}_g(\vec k, z)}{2(P^{ss}_{gg}(\vec k,z) + n_g(z)^{-1})}}\bigg)},
\end{equation}

where $\rm{P^{ss}_{gg}(\vec{k}, z) = b^2_g(k, z) (1 + \beta_g \mu^2{\hat{k}})^2 P_m(k, z)}$ is the three-dimensional power spectrum of galaxies, $\rm{b_g(k, z)}$ is the galaxy bias parameter, $\rm{\beta_g = \frac{f}{b_g}}$ with $\rm{f \equiv \frac{d \ln D}{d \ln a}}$ describing the logarithmic growth function, $\rm{\mu_k = \cos(\hat{n} \cdot \hat{k})}$ is the cosine of the angle between the line of sight $\rm{\hat{n}}$ and the Fourier modes $\rm{\hat{k}}$, and $\rm{n_g(z) = \frac{N_g(z)}{V_s}}$ is the number density of galaxies within the redshift bin $\rm{z}$. The Fourier transform of the galaxy distribution, $\rm{\delta^s_g(\vec{k}, z)}$, is computed as $\rm{\int d^3\vec{r} , \delta_g(\vec{r}) e^{i\vec{k} \cdot \vec{r}}}$. Following the description of the probability distribution and posterior distribution, the likelihood function, $\rm{\mathcal{L}(\vec{\vartheta}_{\text{GW}} | P{gg}(\vec k,z), \Theta_n, \vec d_g(z))}$, is defined in Equation\eqref{eq:LikeliCross}

\begin{widetext}
\begin{equation}
  \mathrm{\mathcal{L} \propto \exp\bigg(-\frac{V_s}{4\pi^2}\int k^2 dk \int d\mu_k
  \frac{ \bigg(\hat{P} (\vec{k}, \Delta \Omega_{GW}) - b_g(k,z)b_{GW}(k, z)(1 + \beta_g \mu_{\hat{k}}^2)P_{m}(k,z)e^{-\frac{k^2}{k^2_{\rm eff}}}\bigg)^2}
  {2\big(P^{ss}_{gg}(\vec{k},z) + n_g(z)^{-1}\big)\big(P^{rr}_{GW\,GW}(\vec{k},z) + n_{GW}(z)^{-1}\big)}
  \bigg),}
\label{eq:LikeliCross}
\end{equation} 
\end{widetext}

where $\rm{\hat{P}(\vec k, z)= \delta_{g}(\vec k, z)\delta_{\text{GW}}^*(\vec k,\Delta \Omega_{\text{GW}})}$ represents the cross power spectrum between the galaxy and GW data. The term $\rm{n_{\text{GW}}(z)= \frac{N_{\text{GW}}(D^i_l(z))}{V_s}}$ is the number density of gravitational wave sources described in terms of the number of objects within a specific luminosity distance bin. $\rm{V_s}$ denotes the total sky volume, and $\rm{P^{rr}_{\text{GW,GW}}(\vec k, z) = b^2_{\text{GW}}(k, z)P_m(k, z)}$ with $\rm{b_{\text{GW}}(k, z)}$ being the GW bias parameter. This formulation for the likelihood function integrates cross power spectrum analysis, crucial for determining the correlation between galaxy distribution and GW sources, with a model that considers the biases and noise properties of both galaxies and gravitational waves. The exponential function captures the discrepancy between the observed and theoretical power spectra, reflecting the statistical nature of the inference process in this studies due to the sky localization error.

\subsection{Redshift Inference Using Cross-Correlation}
\label{sec:methodology}

In this analysis, we utilize catalogs from the two major galaxy surveys: the Large Synoptic Survey Telescope (LSST)\citep{LSSTDarkEnergyScience:2021vfu} simulated catalog and the Dark Energy Spectroscopic Instrument (DESI)\citep{2019AJ....157..168D} simulated catalog. LSST offers photometric measurements and DESI provides spectroscopic measurements. We employ these catalogs to cross-correlate with GW sources to infer redshift information. The LSST catalog, a photometric survey, contains around 3 million galaxies with redshifts up to $\rm{z=3}$ that we use for our study. LSST's wide area coverage and deep field capabilities allow it to observe faint objects, providing data on a diverse range of galaxy types and distances. Its inclusion of galaxies up to high redshifts enables studies of the universe at different epochs. However, LSST relies on photometric redshift estimations, which are less precise than spectroscopic redshifts, potentially introducing errors in redshift determination. The DESI catalog, on the other hand, is a spectroscopic survey designed to measure precise redshifts up to $\rm{z=2}$ with approximately 6 million galaxies. This catalog provides highly accurate redshift measurements, significantly reducing uncertainties in redshift measurements. However, compared to LSST, DESI has a more limited redshift range, potentially missing high-redshift galaxies. 

\begin{figure*}
    \centering
    \subfloat{%
        \includegraphics[height=5.5cm, width=8.5cm]{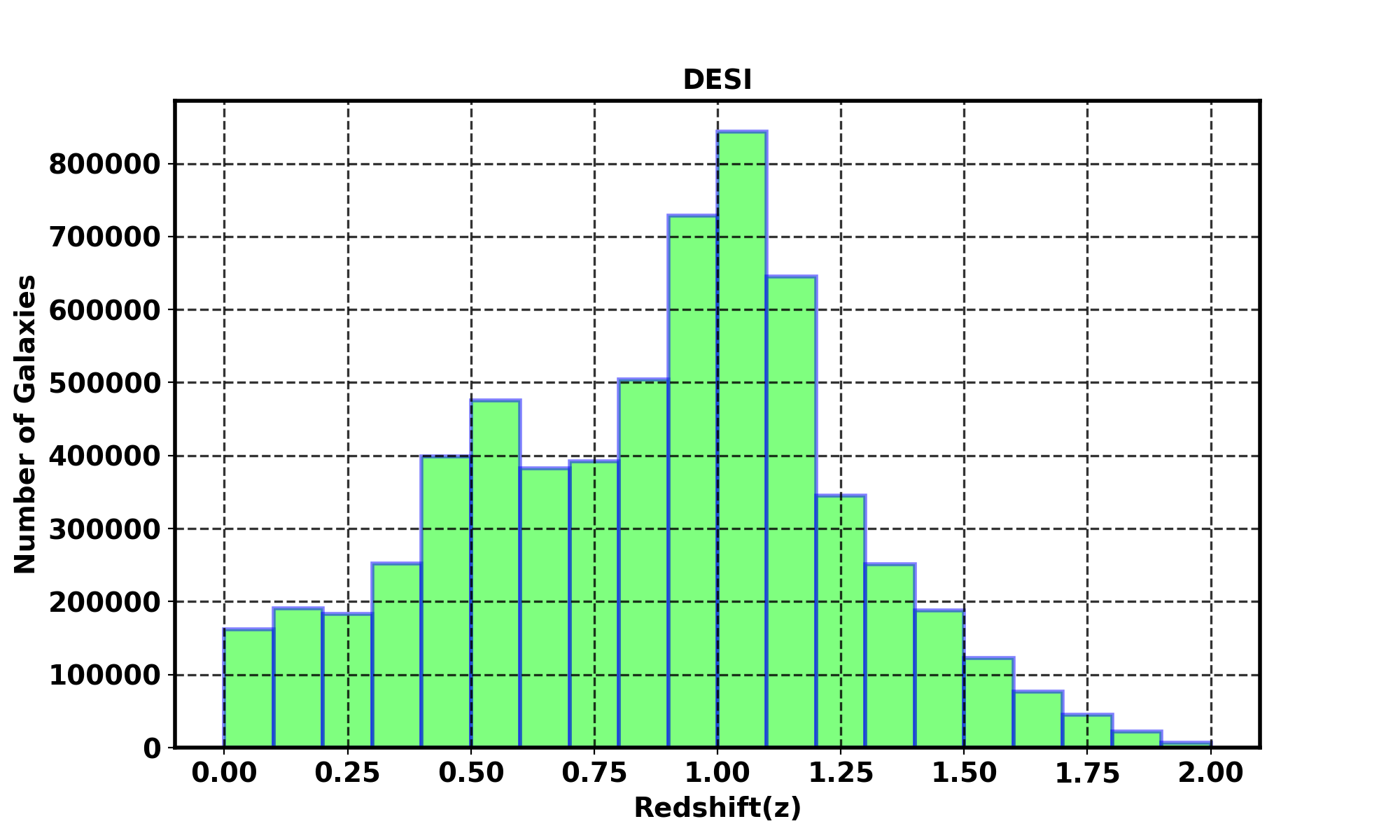}%
        }
    \hfill
    \subfloat{%
        \includegraphics[height=5.5cm, width=8.5cm]{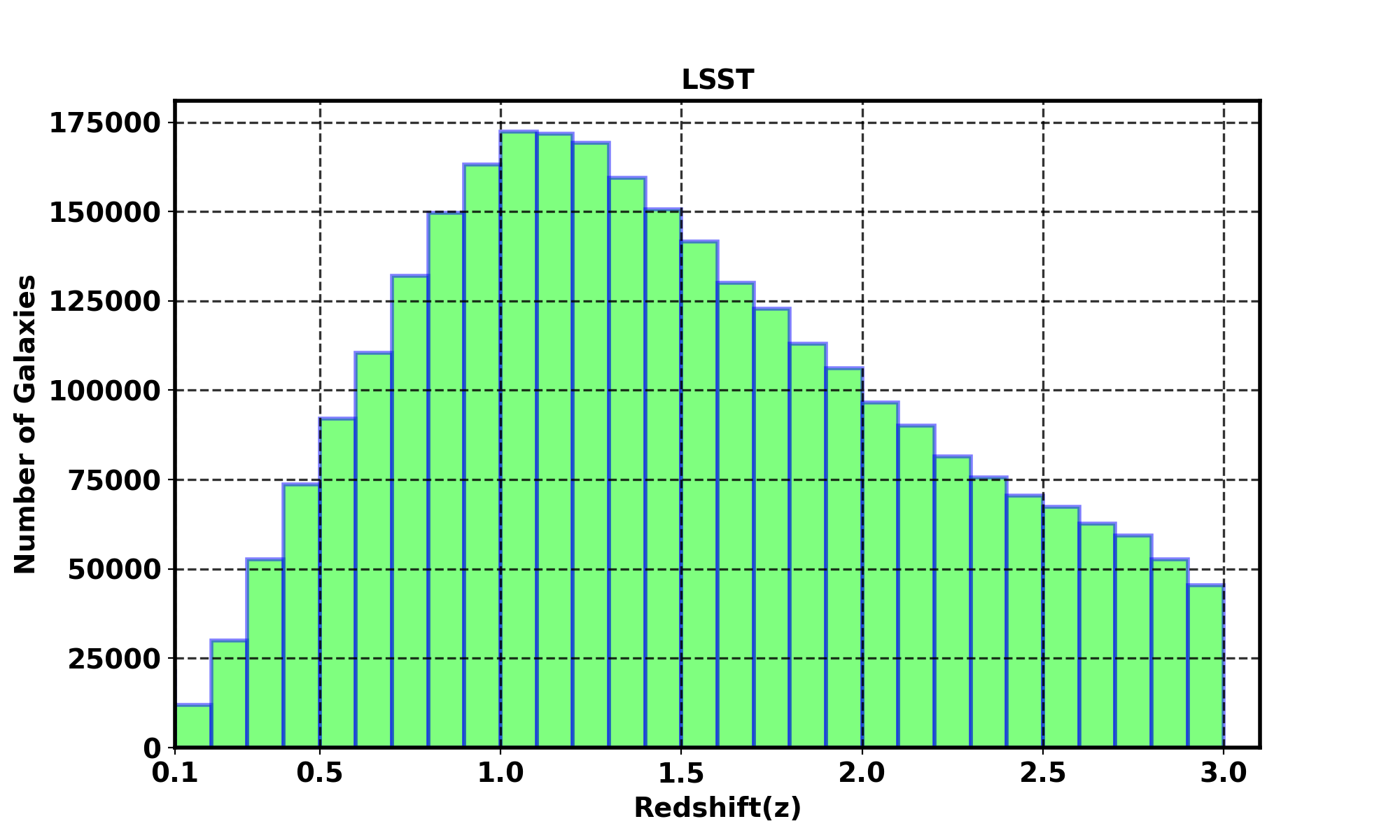}%
        }
    \caption{Histogram illustrating the number of galaxies as a function of redshift for the DESI and LSST catalogs with $\Delta z=0.1$. The DESI catalog contains approximately 6 million galaxies in total, covering a sky fraction of 14,000 square degrees. In contrast, the LSST catalog contains 3 million galaxies because we are using a tiny fraction of the survey in our analysis, although in principle, LSST will cover 18,000 square degrees. The left plot shows the distribution for DESI, and the right plot shows the distribution for the fraction of LSST used in this analysis.}
    \label{fig:catelog}
\end{figure*}

We have taken the DESI and LSST catalogs from the CosmoHub database \citep{2017ehep.confE.488C,Tallada:2020qmg}, which provide a known number of galaxies with measured redshifts ($\rm{z_g}$). We divide these galaxies into tomographic bins with $\rm{N_z}$ galaxies in each redshift bin. The chosen redshift interval ($\rm{\Delta z}$) is 0.05. Similarly, for each redshift bin, the number of GW sources is illustrated in Figure \ref{fig:GWevent}. Neither of these galaxy surveys covers the entire sky. The DESI survey covers approximately 14,000 square degrees, which corresponds to a sky fraction ($\rm{f_{sky}}$) of about 0.339. The LSST survey covers approximately 18,000 square degrees, corresponding to a sky fraction of about 0.436. It is important to note that GW sources are distributed across the full sky, whereas the galaxy surveys only cover part of it. To account for this discrepancy, we assume that the GW sources are homogeneously distributed across the sky. Since the galaxy surveys do not cover the entire sky, we only consider a fraction of the GW sources that corresponds to the fraction of the sky covered by the survey. To generate the GW catalog from the galaxy catalog, we randomly select galaxies from the galaxy catalog with uniform weighting within a tomographic bin structure and assign BBH to it. The number of galaxies and GW sources in each redshift bin are shown in Figures \ref{fig:GWevent} and \ref{fig:catelog}. This process has been applied to both the LSST and DESI surveys. For each redshift bin, we calculate the auto and cross power spectra for the GW catalog and the galaxy catalog using the publicly available package Nbodykit \citep{{Hand:2017pqn}}. We perform these calculations for each redshift bin, starting from $\rm{z = 0}$ to $\rm{z = 2.5}$, using Planck-2015 cosmology parameters \citep{Planck:2015fie}. Additionally, we calculate the matter power spectrum $\rm{P_m(k,z)}$ for each redshift bin using Nbodykit, employing the non linear matter power spectrum. The power spectrum is obtained for the same cosmological parameters used for the calculation of auto and cross power spectra for DESI and LSST. The logarithmic growth function is set to zero in our calculations. The bias parameters for galaxies ($\rm{b_g(k, z)}$) and GW sources ($\rm{b_{GW}(k, z)}$) vary with redshift and scale. At large scales ($\rm{k < 0.1 \, h/\text{Mpc}}$), the galaxy bias is approximately constant, with $\rm{b_g \approx 1.6}$. We expect a similar scale-independent behavior for the GW bias parameter ($\rm{b_{GW}}$) at these large scales, influenced by the large-scale distribution of galaxies. Conversely, at smaller scales ($\rm{k > 0.1 \, h/\text{Mpc}}$), $\rm{b_{GW}}$ becomes scale-dependent due to cluster-scale and galaxy-scale processes such as binary formation, stellar metallicity, and supernovae/AGN feedback. The redshift-dependent bias can be modeled as $\rm{b_{GW}(z) = b_{GW}(1 + z)^\alpha}$. For our purposes, we use $\rm{b_{GW} = 2}$ and $\rm{\alpha = 0}$.

Each galaxy survey has complementary advantages: DESI offers precise spectroscopic redshift measurements but is limited to lower redshifts, while LSST extends observations to higher redshifts with photometric measurements. By cross-correlating these surveys with GW sources, we can infer redshifts for dark sirens. DESI is particularly effective for low-redshift measurements with current GW detectors, whereas LSST's wide coverage and high-redshift reach are better suited for future detectors. This study highlights the complementary roles of DESI and LSST in enhancing redshift inference across different GW detector capabilities.

\begin{figure}
\centering
\includegraphics[height=5.5cm, width=9cm]{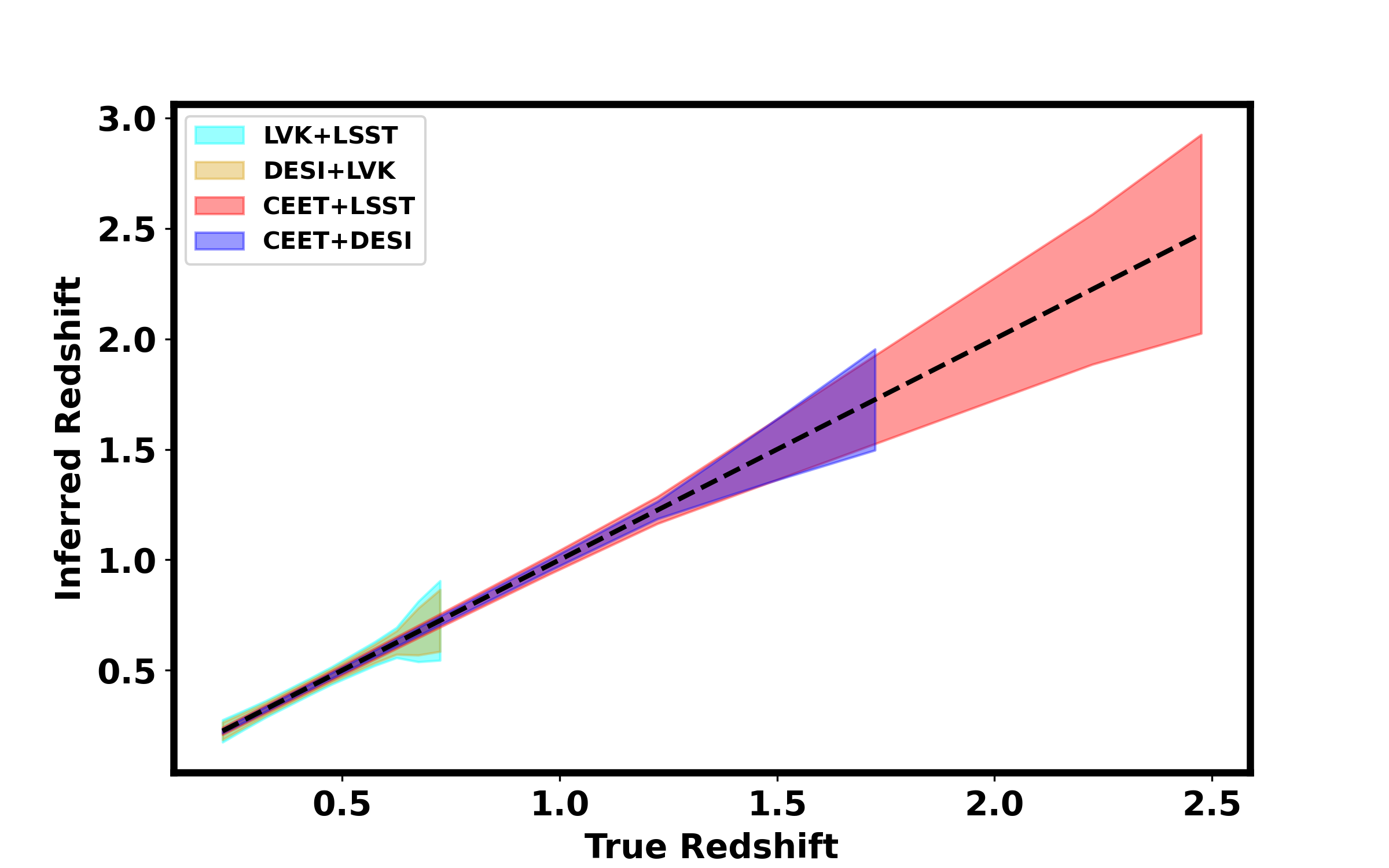}
\caption{This plot shows the errors in inferred redshifts from the cross-correlation between galaxy surveys and GW sources, comparing current detectors, LVK with upcoming ones, CEET. The analysis includes two major galaxy surveys, DESI and LSST, considering both LVK+DESI and LVK+LSST for current detectors, and CEET+DESI and CEET+LSST for the upcoming ones. The results indicate that DESI provides better measurements than LSST, especially at lower redshifts. Additionally, the CEET detectors significantly improve measurement precision compared to the current LVK detectors, highlighting the potential of next-generation detectors to enhance redshift inference accuracy when combined with comprehensive galaxy surveys.}
\label{fig:zerror}
\vspace{-0.1cm}
\end{figure}

Figure \ref{fig:catelog} shows histograms representing the number of galaxies as a function of redshift for the DESI and LSST catalogs. Each histogram bin has a width of $\rm{\Delta z = 0.1}$. The left plot illustrates the distribution for the DESI catalog, which contains approximately 6 million galaxies up to a redshift of 2, covering a sky fraction of 14,000 square degrees. On the right, the distribution for the LSST catalog is displayed, which contains approximately 3 million galaxies because we are using a tiny fraction of the survey in our analysis, although in principle, LSST will cover 18,000 square degrees. These histograms provide a visual representation of the redshift distribution of galaxies in each catalog, offering insights into the population and distribution of galaxies within these surveys.

In Figure \ref{fig:zerror}, we depict the errors in inferred redshifts obtained from the cross-correlation between galaxy surveys and GW sources. This comparison involves current detectors LVK and upcoming ones CEET, combined with two major galaxy surveys, DESI and LSST. The plot includes four scenarios: LVK combined with DESI (LVK+DESI) and LSST (LVK+LSST), as well as CEET combined with DESI (CEET+DESI) and LSST (CEET+LSST). The shaded regions illustrate the error margins for each combination. The results demonstrate that DESI provides superior redshift measurements compared to LSST, particularly at lower redshifts. As the redshift increases, the error margins for both surveys widen. Notably, the upcoming CE and ET detectors offer significantly improved measurement precision over the current LVK detectors, underscoring the potential of next-generation detectors to enhance redshift inference accuracy when integrated with extensive galaxy surveys. The typical error in redshift measurements for LVK with DESI is approximately 6.0\% and for LSST it is approximately 9.0\% within the redshift range of $\rm{z=0.325}$ to $\rm{z=0.625}$. Beyond this range, the error increases significantly. Similarly, for CEET with DESI, the typical error is around 2.8\% and for LSST it is around 3.8\% within the range of 0.275 to 1.2. Errors increase noticeably outside of these ranges for both configurations.

\section{Baryon Acoustic Oscillation Scale from galaxy power spectrum}
\label{sec:BAO}

The phenomenon of BAO emerges from the complex interaction between radiation pressure and gravitational forces in the early Universe. This topic has been extensively studied, with key measurements and theoretical models documented in previous works \citep{peebles1973statistical, crocce2011modelling}. As the Universe cooled and photons decoupled from baryons, a characteristic length scale, known as the sound horizon at the drag epoch ($\rm{z_d}$), referred to as $\rm{r_s}$, left an observable imprint on the large-scale structure of galaxies and the CMB power spectrum \citep{peebles1970primeval, sunyaev1970small, bond1987statistics, hu2002cosmic, blake2003probing}. By analyzing the two-point angular correlation function (2PACF) of galaxy distributions, represented as $\rm{w(\theta)}$, the BAO feature can be detected, appearing as a distinct bump. The BAO scale is determined by fitting the angular correlation function $\rm{w(\theta)}$ using a model that combines a power-law with a Gaussian term as \citep{xu20122, carvalho2016baryon, alam2017clustering}.

\begin{equation}
\mathrm{w(\theta, z)=a_1+a_2\theta^k+a_3exp\biggl(-\frac{(\theta-\theta_{FIT}(z))^2}{\sigma_{\theta}^2}\biggr)},
\label{eq:fit}
\end{equation}
where, the coefficients $\rm{a_i}$ represent the parameters of the model, $\sigma_{\theta}$ denotes the width of the BAO feature, and $\rm{\theta_{FIT}}$ corresponds to the best-fit value of the angular BAO scale. This signature serves as a robust standard ruler, enabling independent measurements of the angular diameter distance denoted as $\rm{D_A(z)}$ \citep{peebles1974statistical, davis1983survey, hewett1982estimation, hamilton1993toward, landy1993bias}. 

\texttt{Modelling of the 2PACF:} Theoretically, one can model the 2PACF $\rm{w(\theta)}$  as 
\begin{equation}
    \mathrm{w(\theta)=\sum_{l\geq 0}\biggl(\frac{2l+1}{4\pi}\biggr)P_l(\cos(\theta))C_l},
\end{equation}
where $\rm{P_{l} (\cos(\theta))}$ is the Legendre polynomial of the $\rm{l^{th}}$ order and $\rm{C_l}$ the angular power spectrum, which can be expressed in terms of the three-dimensional matter power spectrum $\rm{\mathcal{P}(k)}$ as follows \citep{peebles1973statistical, crocce2011modelling}
\begin{equation}
\mathrm{C_l=\frac{1}{2\pi^2}\int4\pi k^2\mathcal{P}(k)\psi_l^2(k)e^{-k^2/k_{char}^2}},
\end{equation}
where the quantity $\rm{\psi_l(k)}$ is defined as:  
\begin{equation}
    \mathrm{\psi_l(k)=\int dz\phi(z)j_l(kr(z))},
    \label{eq:selection}
\end{equation}
where $\rm{\phi(z)}$ is the galaxy selection function and $j_l$ is the spherical Bessel function of the $\rm{l^{th}}$ order, and $\rm{r(z)}$ is the comoving distance.
To improve the convergence of the integral for the angular power spectrum, a damping factor $\rm{e^{(-k^2/k_{\rm char}^2)}}$ is introduced \citep{anderson2012clustering}. Throughout the calculations, we adopt a fixed value of $\rm{1/k_{\rm char}}$ = 3 Mpc/h, which has no significant impact on the angular power spectrum for the scales of interest. 

\begin{figure}
\centering
\includegraphics[height=5.5cm, width=9.0cm]{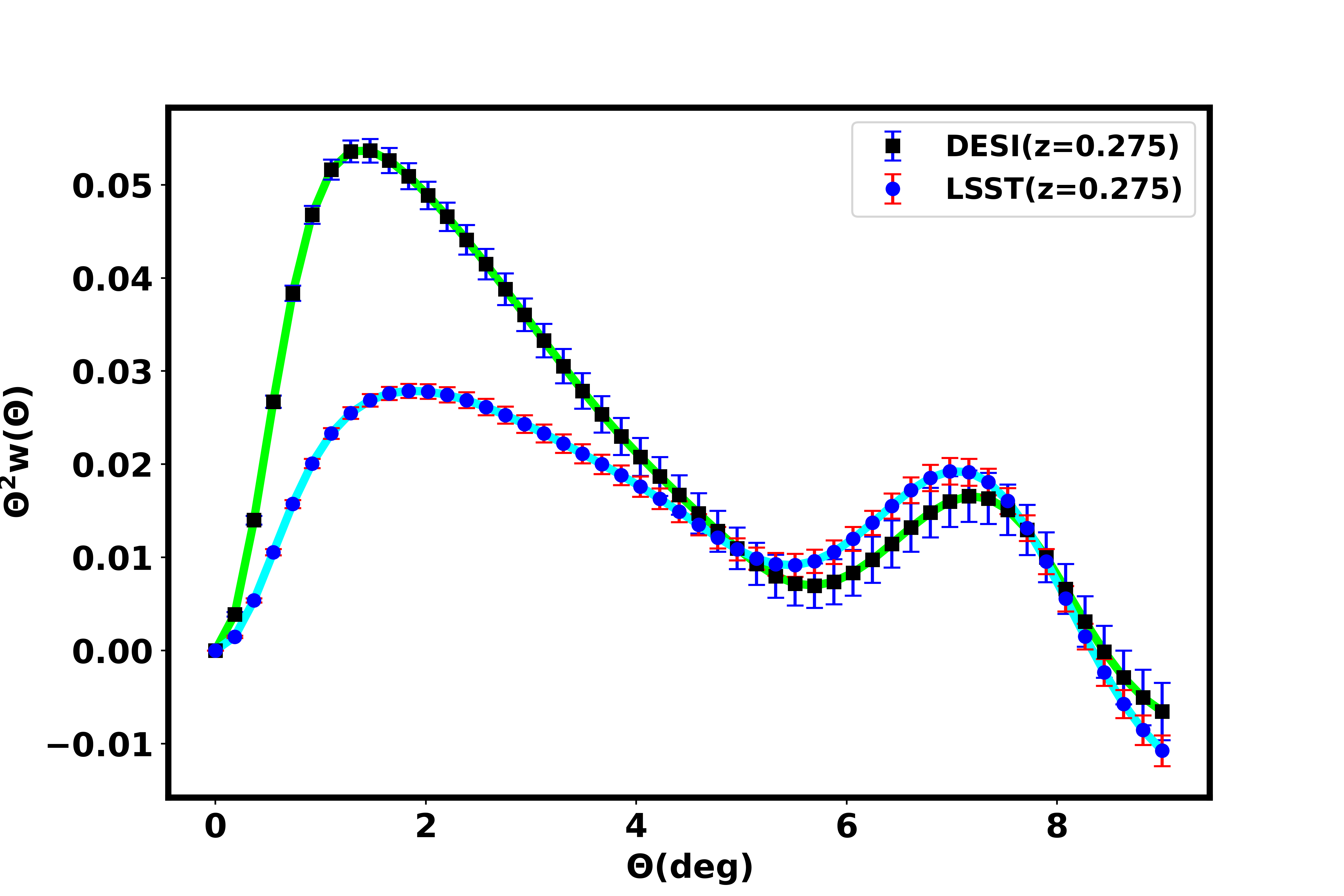}
\caption{This plot illustrates the 2PACF ($\rm{\theta^2 w(\theta)}$), and the angular separation $\theta$ for a specific redshift, $\rm{z = 0.275}$. The calculations are based on the non-linear matter power spectrum for both DESI and LSST. Notably, there is a distinct bump around $\rm{\theta \approx 7.0}$, indicating the presence of the BAO characteristic scale, $\rm{\theta_{\text{BAO}}}$. Additionally, the plot shows that the error bars for DESI are smaller than those for LSST, as expected.}
\label{fig:bao}
\vspace{-0.5cm}
\end{figure}

In this study, we calculate the non-linear matter power spectrum from the module Nbodykit \citep{Hand:2017pqn} , and the selection function on redshift is formulated as a normalized Gaussian function. The standard deviation of this Gaussian function is set to be $\rm{0.03(1+z)}$ for LSST and $\rm{0.01(1+z)}$ for DESI of the mean value.

Several observatories are set to measure the BAO scale with remarkable precision, exploring the universe's redshift depths. Among these instruments, DESI \citep{collaboration2023early} is notable as a pioneering 5-year ground-based experiment designed to study BAO and the evolution of cosmic structures through redshift-space distortions. Spanning a substantial area of 14,000 square degrees in the sky, DESI aims to achieve highly accurate measurements of the BAO feature. The objective is to attain a precision of 0.5\%, targeting the redshift range $\rm{0.0 < z < 3.7}$ \citep{collaboration2023early}. Additionally, Euclid \citep{laureijs2011euclid} and the Vera Rubin Observatory \citep{abell2009lsst, ivezic2019lsst}, each covering an extensive sky area of 18,000 square degrees, will also be capable of making precise BAO measurements up to high redshifts. Figure \ref{fig:bao} illustrates how the quantity $\rm{\theta^2w(\theta)}$ varies with angular separation at a specific redshift $\rm{z=0.275}$. The notable peak at \(\theta \sim 7.0\) is a key observation, highlighting the characteristic scale associated with BAO, which is essential for our study. These calculations are based on the non-linear matter power spectrum for both DESI and LSST surveys, incorporating the selection functions specific to each survey as defined in Equation \eqref{eq:selection}. As expected, the BAO feature is more pronounced for DESI than for LSST, reflecting the higher precision of the spectroscopic survey compared to the photometric one. This plot effectively illustrates the differences in observational capabilities and their impact on BAO feature detection between the two surveys. The covariance of $\rm{w(\theta)}$, denoted as $\rm{Cov_{\theta\theta'}}$, is modeled accordingly.

\begin{figure*}
    \centering
    \subfloat{
        \includegraphics[height=5.5cm, width=9.0cm]{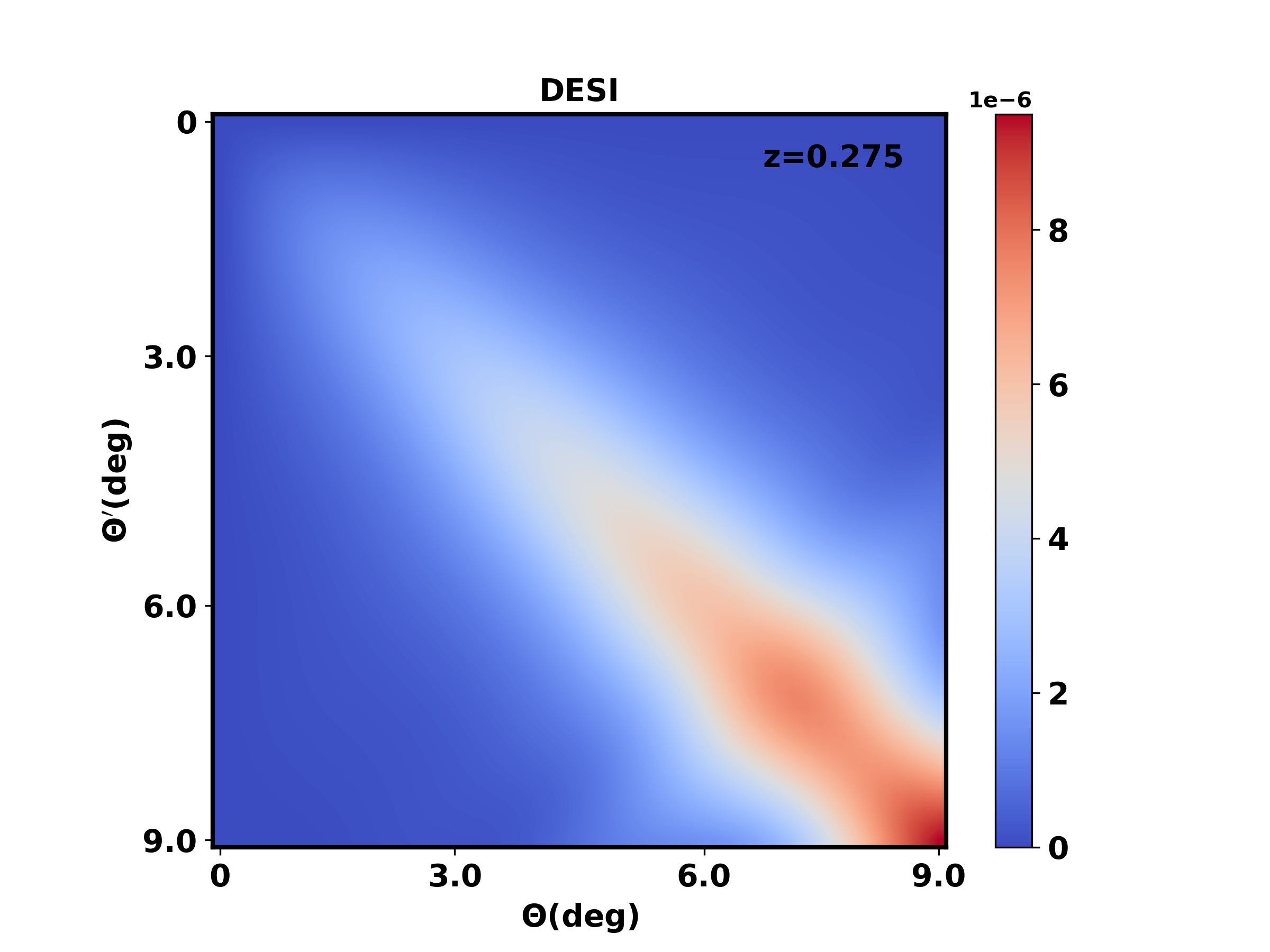}%
        }
    \hfill
    \subfloat{
        \includegraphics[height=5.5cm, width=8.7cm]{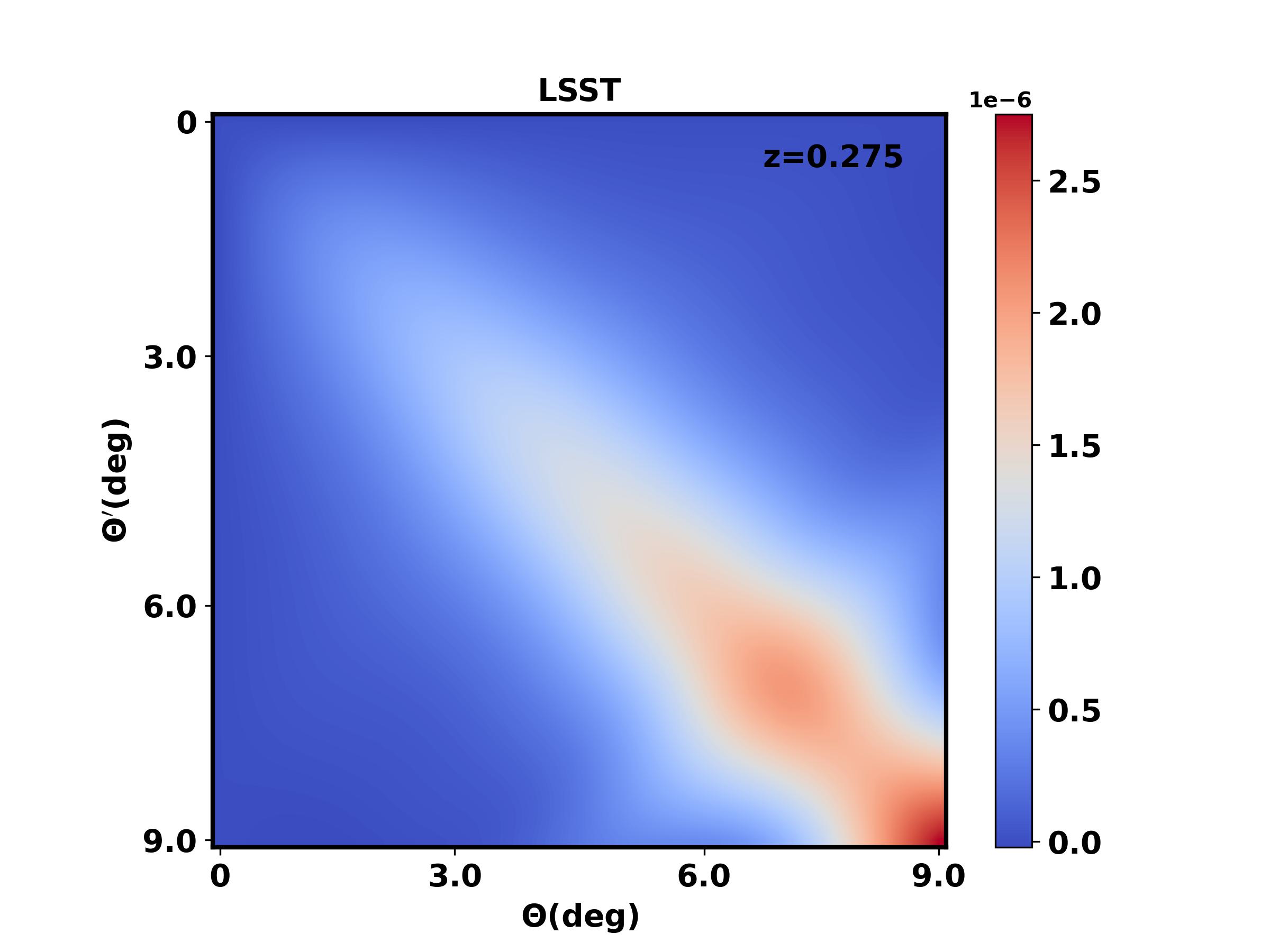}%
        }
    \caption{This plot displays the covariance matrices ($\rm{Cov_{\theta\theta'}}$) calculated using the non-linear matter power spectrum as a function of $\rm{\theta}$ and $\rm{\theta'}$ for a specific redshift, $\rm{z = 0.275}$. It is evident that as both $\rm{\theta}$ and $\rm{\theta'}$ increase, the value of the covariance matrix also increases, indicating larger errors at the larger angular scales. The left plot corresponds to the DESI data, and the right plot corresponds to the LSST survey.}
    \label{fig:covmat}
\end{figure*}

\begin{figure}
\centering
\includegraphics[height=5.5cm, width=9cm]{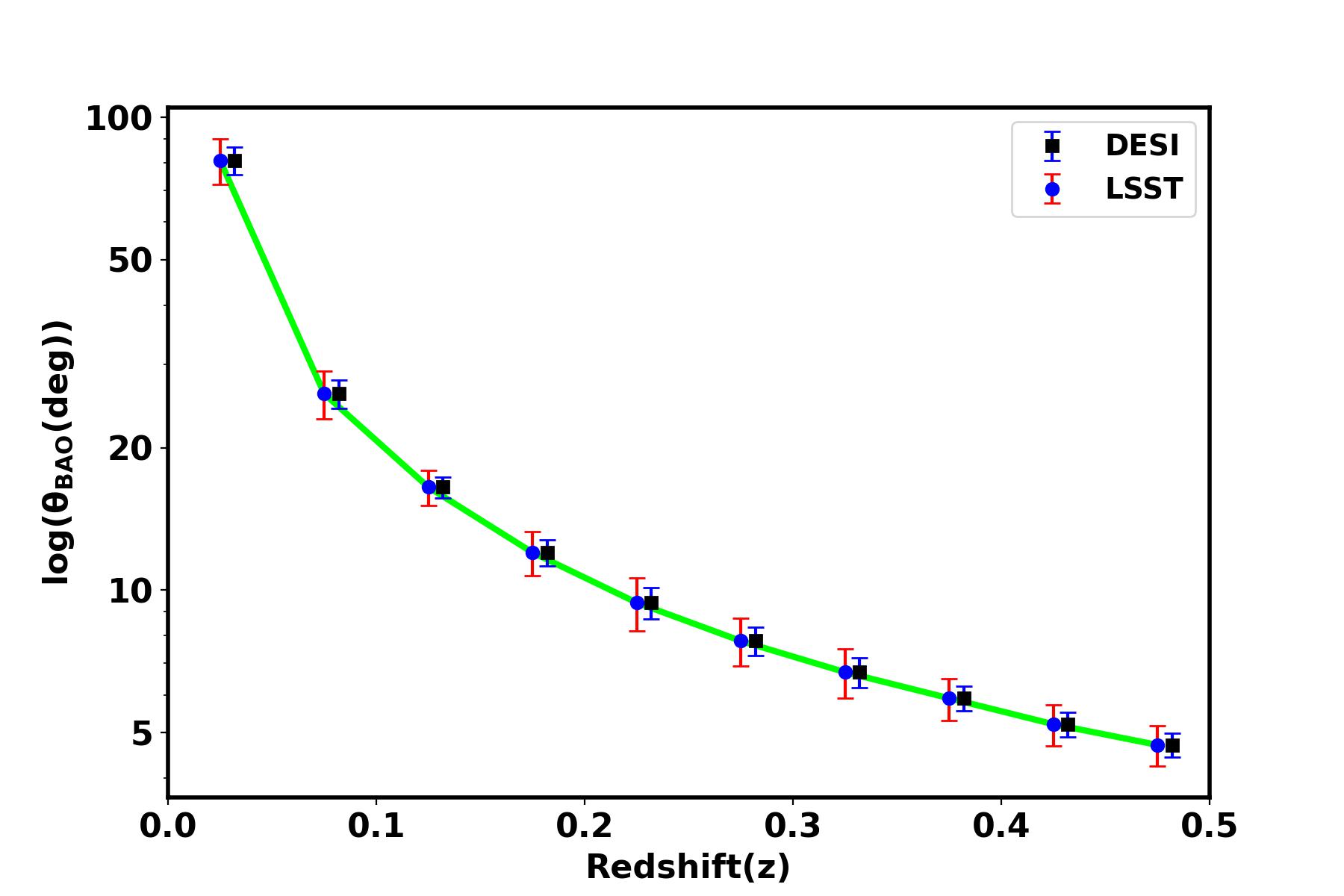}
\caption{We depict the relationship between the BAO scale and its associated error, primarily arising from redshift inaccuracies, for both the LSST and DESI surveys as a function of redshift ($\rm{z}$) up to $\rm{z = 0.5}$. This presentation provides crucial insights into how the BAO scale and its error vary with $\rm{z}$. For visual clarity, the error bars for DESI are shifted slightly to the right. This also demonstrates that the uncertainty in the BAO measurements is larger for LSST than for DESI, which is to be expected.}
\label{fig:baoplot}
\vspace{-0.1cm}
\end{figure}
\begin{equation}
    \mathrm{Cov_{\theta\theta'}=\frac{2}{f_{sky}}\sum_{l\geq 0}\frac{2l+1}{(4\pi)^2}P_l(cos(\theta)) P_l(cos(\theta'))\biggl(C_l+\frac{1}{n}\biggr)^2},
\end{equation}

Here, $\rm{1/n}$ represents the shot noise, which is associated with the number density of galaxies $n$ or, more specifically, the number of objects per steradian. Additionally, $\rm{f_{sky}}$ indicates the fraction of the sky covered by the survey or observation \citep{cabre2007error, dodelson2020modern}. Figure \ref{fig:covmat} shows how the covariance matrix, derived from the non-linear matter power spectrum, varies with angular separations at a specific redshift. The increase in the covariance matrix as both $\rm{\theta}$ and $\rm{\theta'}$ increase suggests a correlation between these angular scales. Figure \ref{fig:baoplot} offers a detailed view of how the BAO scale and its associated error (calculated using Equation \eqref{eq:fit}) evolve with redshift for both DESI and LSST, highlighting the impact of redshift inaccuracies. This also illustrates that the uncertainty in BAO measurements is greater for LSST than for DESI, as anticipated.

\section{Reconstructing F(z) with Redshift: Forecasts \& Results}
\label{sec:Forecast}

In this study, we investigate the measurements of the frictional term $\rm{\mathcal{F}(z)}$ under two distinct scenarios: one with a fixed Hubble constant ($\rm{H_0}$) and another with a varying $\rm{H_0}$. The inclusion of $\rm{H_0}$ measurements in our analysis is essential due to its potential to introduce bias in $\rm{\mathcal{F}(z)}$ measurements. Furthermore, the incorporation of $\rm{H_0}$ enhances the refinement of our results and introduces an additional dimension, which is instrumental in addressing the Hubble tension problem. This problem arises from the notable discordance between the locally measured value of the Hubble constant in the nearby universe and the value inferred from the early universe using the CMB. 

\subsection{Case with the fixed Hubble constant}\label{sec:fzfh0}

The intricate connection between the EM wave luminosity distance at a redshift ($\rm{z}$) and key cosmological parameters, including the BAO scale ($\theta_{BAO}$) and the sound horizon ($\rm{r_s}$) is elegantly encapsulated by the equation
\begin{equation}
    \mathrm{\mathcal{F}(z) = \frac{D_l^{\text{GW}}(z) \theta_{\text{BAO}}(z)}{(1+z) r_s}}.
    \label{eq:fzeqn}
\end{equation}
This equation serves as the cornerstone for the analysis of the frictional component in cosmological studies, correlating three distinct length scales at a redshift ($\rm{z}$). The measurement of redshift utilizes the cross-correlation technique between GW sources and galaxy surveys. The GW luminosity distance ($\rm{D_l^{GW}}$) for detectable GW events is determined using \texttt{Bilby}. The BAO scale \(\rm{(\theta_{\text{BAO}})}\) is derived from the 2PACF using Equation \eqref{eq:fit}. This method is versatile, allowing for application across various surveys without being restricted to any specific one. The sound horizon is measured through the analysis of acoustic oscillations in the CMB radiation, employing several missions (such as WMAP \citep{hinshaw2013nine}, Planck \citep{aghanim2020planck}, ACTPol \citep{thornton2016atacama}, and SPT-3G \citep{sobrin2022design}). The first peak in the angular power spectrum of the CMB corresponds to the sound horizon scale at the time of recombination. The sound horizon value at the drag epoch, measured with high precision by Planck, is \(\rm{147.09 \pm 0.26}\) Mpc \citep{aghanim2020planck}.

We employ a hierarchical Bayesian framework to infer the posterior distribution of the frictional function, $\rm{\mathcal{F}(z)}$, as a function of redshift for a total of $\rm{n_{{GW}}}$ GW sources. The posterior distribution is given by the equation

\begin{equation}
\begin{aligned}
    &\mathrm{P(\mathcal{F}(z)) \propto \Pi(\mathcal{F}(z)) \prod_{i=1}^{n_{\mathrm{GW}}} \iiiint dr_s \, dD_l^{\mathrm{GW}^i} \, d\theta_{\mathrm{BAO}}(z^i)}  \\ \times & \mathrm{dz^i \, P(r_s) P(\theta_{\mathrm{BAO}}(z^i)) P(z^i) \mathcal{L}(D_l^{\mathrm{GW}^i} | \mathcal{F}(z^i), \theta_{\mathrm{BAO}}(z^i), r_s, z^i),}
\end{aligned}
\label{eq:FZposterior}
\end{equation}

here, $\rm{P(\mathcal{F}(z))}$ denotes the posterior probability density function of $\rm{\mathcal{F}(z)}$, given the data from GW  sources $\rm{\{D_l^{\mathrm{GW}}\}}$ and their redshifts $\rm{z}$, obtained using cross-correlation techniques between GW sources and galaxy surveys. The measurements of the BAO scale, $\rm{\theta_{\mathrm{BAO}}}$, at the redshift $\rm{z}$ of the GW source, are acquired from the galaxy surveys. The likelihood function is represented by $\rm{\mathcal{L}(D_l^{\mathrm{GW}^i} | \mathcal{F}(z^i), \theta_{\mathrm{BAO}}(z^i), r_s, z^i)}$. The terms $\rm{P(r_s), P(z^i)}$, and $\rm{P(\theta_{\mathrm{BAO}}(z^i))}$ denote the prior probability distributions for the sound horizon $\rm{r_s}$, redshift $\rm{z}$, and BAO scale, respectively. The term $\rm{\Pi(\mathcal{F}(z))}$ represents the prior distribution for $\rm{\mathcal{F}(z)}$, which reflects our prior knowledge or assumptions about the frictional term before incorporating the measured data. A flat prior on $\rm{\mathcal{F}(z)}$ ranging from 0.1 to 2 was chosen to ensure a non-informative prior that covers a broad range of possible deviations from GR while avoiding extreme values that are physically implausible or poorly constrained by the data. This range allows us to explore a wide parameter space without imposing strong biases on the results.

\subsection{Reconstruction of F(z)}
\label{sec:recfz}
\begin{figure*}
\centering
\includegraphics[height=7.0cm, width=18cm]{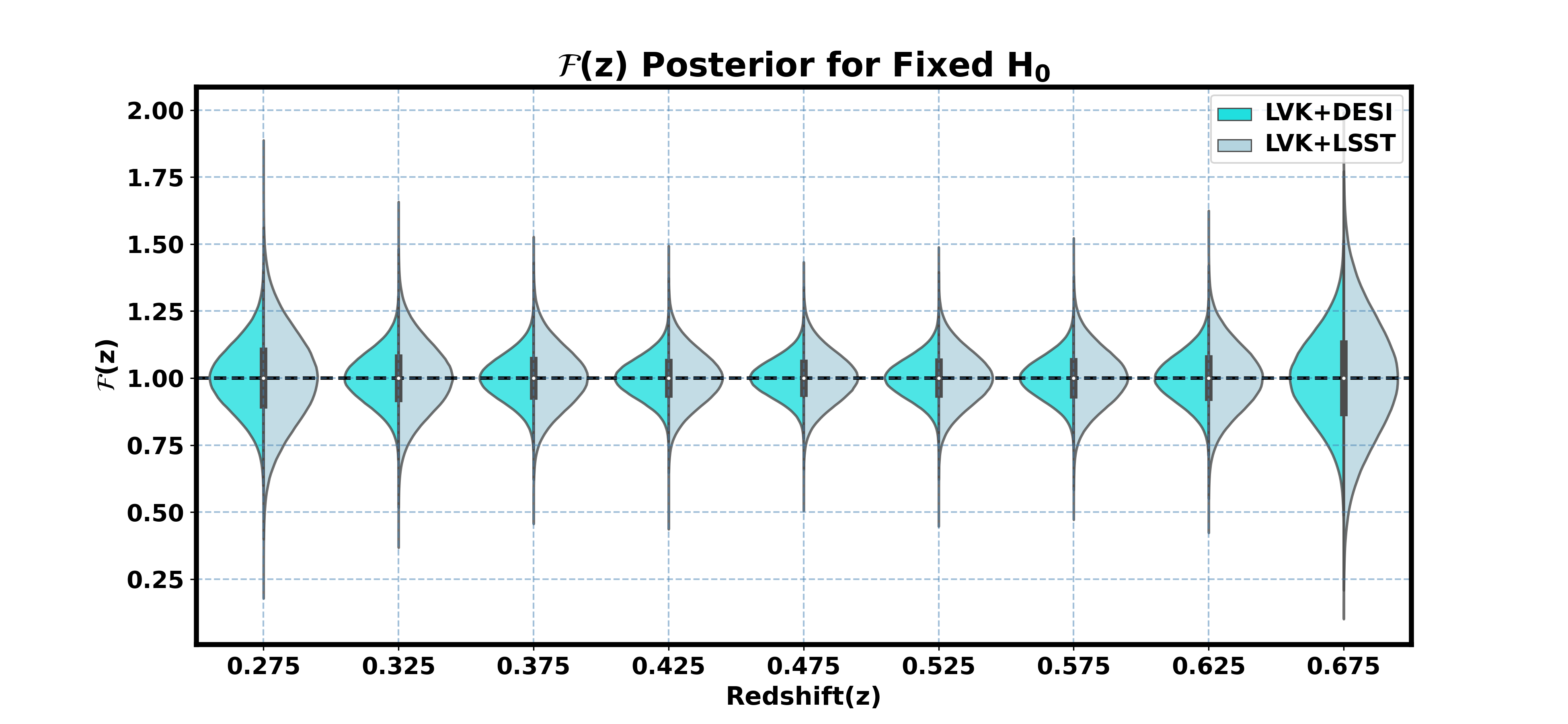}
\caption{
This violin plot illustrates the posterior on the reconstruction of the non-GR parameter $\rm{\mathcal{F}(z)}$ as a function of cosmic redshift for the LVK system, considering both the galaxy survey DESI (LVK+DESI) and LSST (LVK+LSST), for binary black hole mergers up to a redshift of $\rm{z=0.675}$. The plot presents results under fixed Hubble constant scenarios. This plot highlights the potential for achieving more accurate measurements of $\rm{\mathcal{F}(z)}$ at significantly deeper cosmic redshifts for current GW detectors when combined with galaxy surveys like DESI and LSST. Additionally, it suggests that DESI provides better measurements than LSST, as expected at lower redshifts.}
\label{fig:LVKviolin}
\end{figure*}

\begin{figure*}
\centering
\includegraphics[height=7.0cm, width=18cm]{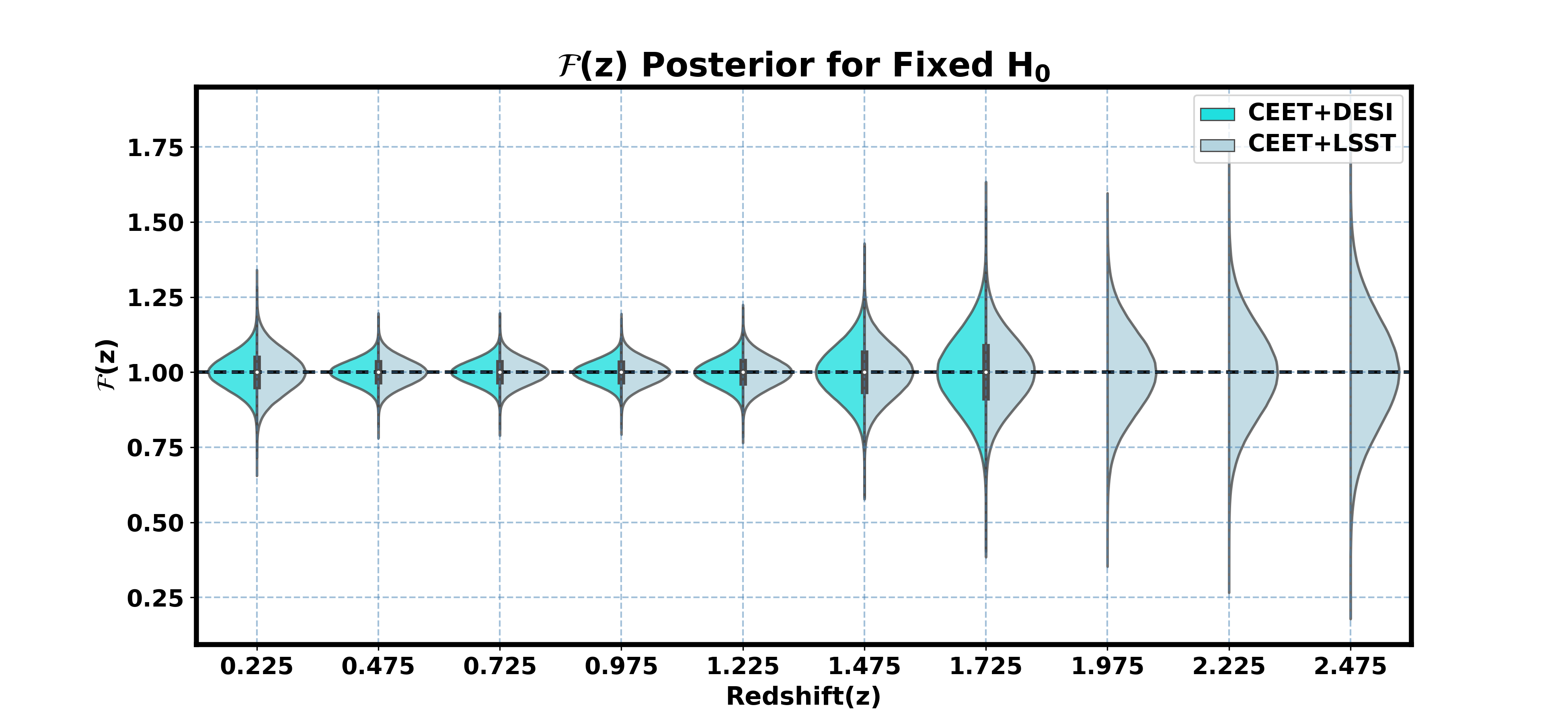}
\caption{This violin plot illustrates the posterior on the reconstruction of the non-GR parameter $\rm{\mathcal{F}(z)}$ as a function of cosmic redshift for CE and ET systems, considering both the galaxy survey DESI (CEET+DESI) and LSST (CEET+LSST), for binary black hole mergers up to a redshift of $\rm{z=2.5}$. The plot presents results under fixed Hubble constant scenarios. This plot highlights the potential for achieving more accurate measurements of $\rm{\mathcal{F}(z)}$ at significantly deeper cosmic redshifts for future GW detectors like CE and ET when combined with galaxy surveys like DESI and LSST. It also indicates that DESI gives better measurements at low redshifts, while LSST provides better measurements at high redshifts. This is expected because DESI is a spectroscopic survey, which excels at low redshifts where individual galaxy spectra can be obtained for precise redshift measurements. In contrast, LSST is a photometric survey, which is more effective at higher redshifts where obtaining individual galaxy spectra becomes more challenging.}
\label{fig:CEETviolin}
\end{figure*}

This study focuses exclusively on dark sirens, presenting cases for both the CEET and LVK systems. For both systems, the local merger rate ($\rm{R_0}$) is set at 20 $\rm{Gpc^{-3}yr^{-1}}$ with a 75\% duty cycle. 
In figure \ref{fig:GWevent} presents the number of detectable events as a function of redshift for both the LVK and CEET systems. Using these sources, we show in figures \ref{fig:LVKviolin} and \ref{fig:CEETviolin} a comprehensive comparison of $\mathcal{F}(z)$ measurements under the scenario of fixed  $H_0$ for both the LVK and CEET sources respectively by combining with DESI and LSST. This comparative analysis highlights the impact of galaxy surveys on the precision of $\rm{\mathcal{F}(z)}$ measurements. Additionally, it underscores the potential of advanced detectors like CEET to deliver increasingly accurate cosmological insights, extending deep into higher redshifts. 

The error associated with the frictional term is primarily dominated by the redshift error inferred from the cross-correlation of galaxy surveys and GW sources. Other sources of error include the BAO scale from large-scale surveys, sound horizon distance measurements from the CMB, and from the GW side. GW-related errors stem from two main sources: sky localization error (discussed in Section \ref{sec:RedshiftInf}) and luminosity distance error ($\rm{D_l^{GW}}$). The accuracy of luminosity distance measurements in GW astronomy is influenced not only by the sensitivity and configuration of the detector (see Section \ref{sec:ParamEstBilby}) but also significantly by weak lensing, especially for sources detected at high redshifts by CEET. Gravitational lensing affects GW similarly to EM radiation. For GW events expected from large redshifts ($\rm{z > 1}$), weak lensing is common, along with occasional strongly-lensed sources. A lens with magnification $\rm{\mu}$ modifies the observed luminosity distance to $\rm{\frac{D_l^{GW}}{\sqrt{\mu}}}$, introducing a systematic error $\rm{\Delta D_l^{GW} / D_l^{GW} = 1 - \frac{1}{\sqrt{\mu}}}$ \citep{Mpetha:2022xqo}. This lensing error, when convolved with the expected magnification distribution $\rm{p(\mu)}$ from a standard $\Lambda$CDM model, significantly influences the overall measurement accuracy. In this study, we use the following model to estimate the error due to weak lensing \citep{Hirata:2010ba} 
\begin{equation}
    \mathrm{\frac{\sigma_{WL}}{D_l^{GW}} = \frac{0.096}{2} \left(\frac{1 - (1 + z)^{-0.62}}{0.62}\right)^{2.36}}.
\end{equation}
The total uncertainty in the measured luminosity distance is therefore expressed as $\mathrm{\sigma_{D_l^{GW}}^2 = \sigma_{GW}^2 + \sigma_{WL}^2}$, where $\rm{\sigma_{GW}}$ is the uncertainty derived from the detector's setup during parameter estimation, and $\rm{\sigma_{WL}}$ accounts for the lensing effects. This comprehensive approach ensures a more accurate understanding of the intrinsic properties of GW sources.

In this section, we measure $\rm{\mathcal{F}(z)}$ using a constant $\rm{H_0}$ that remains consistent across redshifts, acting as an overall normalization factor for $\rm{\mathcal{F}(z)}$ measurements. This means that if we can obtain a precise and independent measurement of $\rm{\mathcal{F}(z)}$ in our local universe, we can normalize it to 1, thereby removing any dependence on $\rm{H_0}$. This normalization is particularly effective at low redshifts, where the cumulative effects of modified gravity wave propagation have not yet caused significant deviations from GR. The typical error on the measurements of $\rm{\mathcal{F}(z)}$ with a fixed $\rm{H_0}$ scenario for LVK+DESI ranges around 9\% between $\rm{z=0.325}$ and $\rm{z=0.625}$, while for LVK+LSST it is around 12\%. Outside this range, the error increases significantly, as shown in Figure \ref{fig:LVKviolin}. Similarly, for CEET+DESI, the typical error is around 4\% between $\rm{z=0.275}$ and $\rm{z=1.2}$, and for CEET+LSST, it is around 5\%. Beyond these ranges, the error also increases significantly, as depicted in Figure \ref{fig:CEETviolin}.

\subsection{Case with varying the Hubble constant}\label{sec:fzvh0}

In the preceding sections, our analysis focused on evaluating the frictional term $\rm{\mathcal{F}(z)}$ from Equation \eqref{eq:fzeqn}, assuming a fixed Hubble constant ($\rm{H_0}$). However, in this section, we extend our analysis to jointly infer both $\rm{\mathcal{F}(z)}$ and $\rm{H_0}$. This approach treats $\rm{H_0}$ as a free parameter, serving as an overall normalization for the $\rm{\mathcal{F}(z)}$ measurements based on the sound horizon, which remains constant across the relevant redshifts.

Our framework employs a hierarchical Bayesian analysis to jointly estimate $\rm{\mathcal{F}(z)}$ and $\rm{H_0}$, effectively handling their unique uncertainties and systematics. The resulting correlated constraints significantly contribute to our understanding of cosmological phenomena, addressing potential discrepancies between datasets and revealing insightful connections among fundamental cosmological parameters. The equation representing the joint posterior distribution on $\rm{\mathcal{F}(z)}$ and $\rm{H_0}$ for these sources is given as in Equation \eqref{eq:FZH0posterior}.

\begin{widetext}
\begin{equation}
    \mathrm{P(\mathcal{F}(z), H_0) \propto \Pi(\mathcal{F}(z)) \Pi(H_0) \prod_{i=1}^{n_{\mathrm{GW}}} \iiiint dr_s \, dD_l^{\mathrm{GW}^i} \, d\theta_{\mathrm{BAO}}(z^i) \, dz^i P(r_s) P(\theta_{\mathrm{BAO}}(z^i)) P(z^i) \mathcal{L}(D_l^{\mathrm{GW}^i} | \mathcal{F}(z^i), H_0, \theta_{\mathrm{BAO}}(z^i), r_s, z^i).}
\label{eq:FZH0posterior}
\end{equation}
\end{widetext}

In this equation, $\rm{P(\mathcal{F}(z), H_0)}$ denotes the joint posterior probability density function of $\rm{\mathcal{F}(z)}$ and $\rm{H_0}$. The term $\rm{\mathcal{L}(D_l^{GW^i}|\mathcal{F}(z)^i, H_0^i, \theta_{BAO}(z^i), r_s,z^i)}$ corresponds to the likelihood function. The term $\rm{\Pi(H_0)}$ denotes the prior distribution for $\rm{H_0}$, encapsulating our knowledge or assumptions regarding the Hubble constant before incorporating measured data. All other terms represent the same entities as defined in Section \ref{sec:fzfh0}. In our hierarchical Bayesian framework, we adopt a flat prior assumption, indicating a lack of prior knowledge about the parameters for both the frictional term ($\mathcal{F}(z)$) and the cosmic expansion rate ($\rm{H_0}$). The parameter ranges are defined from 0.1 to 2.0 on $\rm{\mathcal{F}(z)}$ and from 20 km/s/Mpc to 120 km/s/Mpc on $\rm{H_0}$. The choice of prior $\rm{\mathcal{F}(z)}$ is taken large enough to ensure that it can capture  theoretical predictions of beyond GR models\citep{deffayet2007probing, saltas2014anisotropic, nishizawa2018generalized, belgacem2018gravitational, belgacem2018modified, lombriser2016breaking, lombriser2017challenges}.

\subsection{Reconstruction of F(z) and Hubble Constant}
\label{sec:ReconofH0andFz}

\begin{figure*}
\centering
\includegraphics[height=7.0cm, width=18cm]{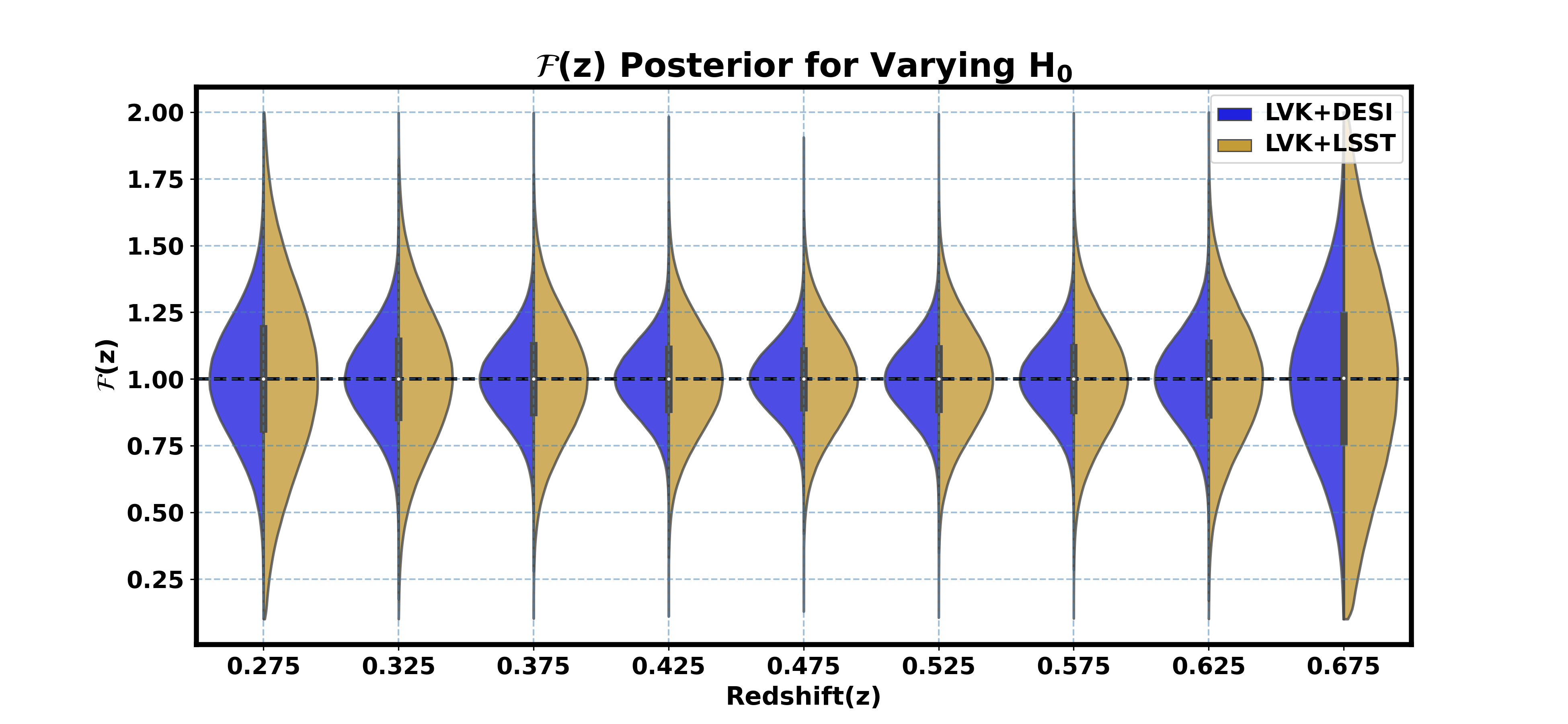}
\caption{
This violin plot illustrates the posterior on the reconstruction of the non-GR parameter $\rm{\mathcal{F}(z)}$ as a function of cosmic redshift for the LVK system, considering both the galaxy survey DESI (LVK+DESI) and LSST (LVK+LSST), for binary black hole mergers up to a redshift of $\rm{z=0.675}$. The plot presents results under varying Hubble constant scenarios. This plot highlights the potential for achieving more accurate measurements of $\rm{\mathcal{F}(z)}$ at significantly deeper cosmic redshifts for current GW detectors when combined with galaxy surveys like DESI and LSST. Additionally, it suggests that DESI provides better measurements than LSST, as expected at lower redshifts.}
\label{fig:LVKviolinH0}
\end{figure*}

\begin{figure*}
\centering
\includegraphics[height=7.0cm, width=18cm]{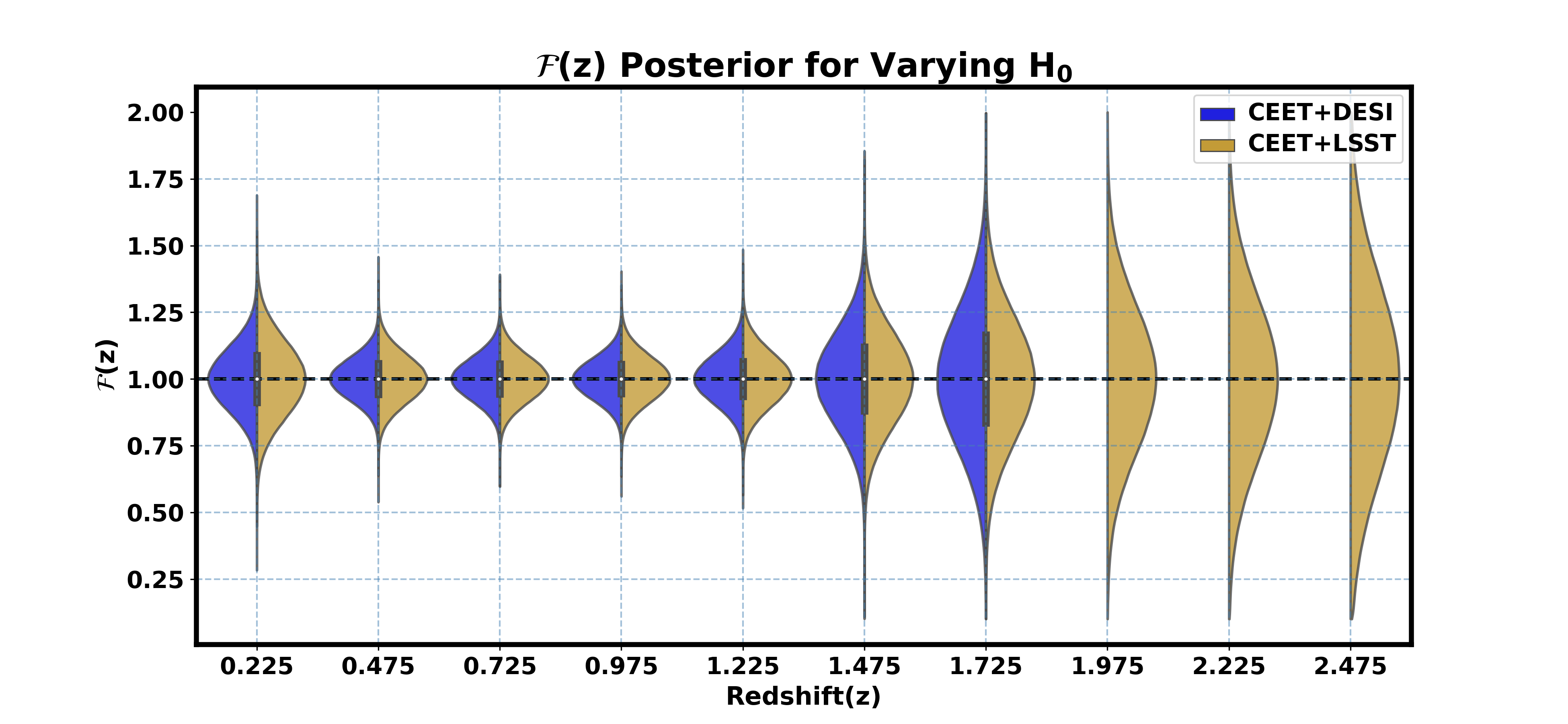}
\caption{This violin plot illustrates the posterior on the reconstruction of the non-GR parameter $\rm{\mathcal{F}(z)}$ as a function of cosmic redshift for CE and ET systems, considering both the galaxy survey DESI (CEET+DESI) and LSST (CEET+LSST), for binary black hole mergers up to a redshift of 2.5. The plot presents results under varying Hubble constant scenarios. This plot highlights the potential for achieving more accurate measurements of $\rm{\mathcal{F}(z)}$ at significantly deeper cosmic redshifts for future GW detectors like CE and ET when combined with galaxy surveys like DESI and LSST. It also indicates that DESI gives better measurements at low redshifts, while LSST provides better measurements at high redshifts. This is expected because DESI is a spectroscopic survey, which excels at low redshifts where individual galaxy spectra can be obtained for precise redshift measurements. In contrast, LSST is a photometric survey, which is more effective at higher redshifts where obtaining individual galaxy spectra becomes more challenging.}
\label{fig:CEETviolinH0}
\end{figure*}

\begin{figure}
\centering
\includegraphics[height=6.0cm, width=9cm]{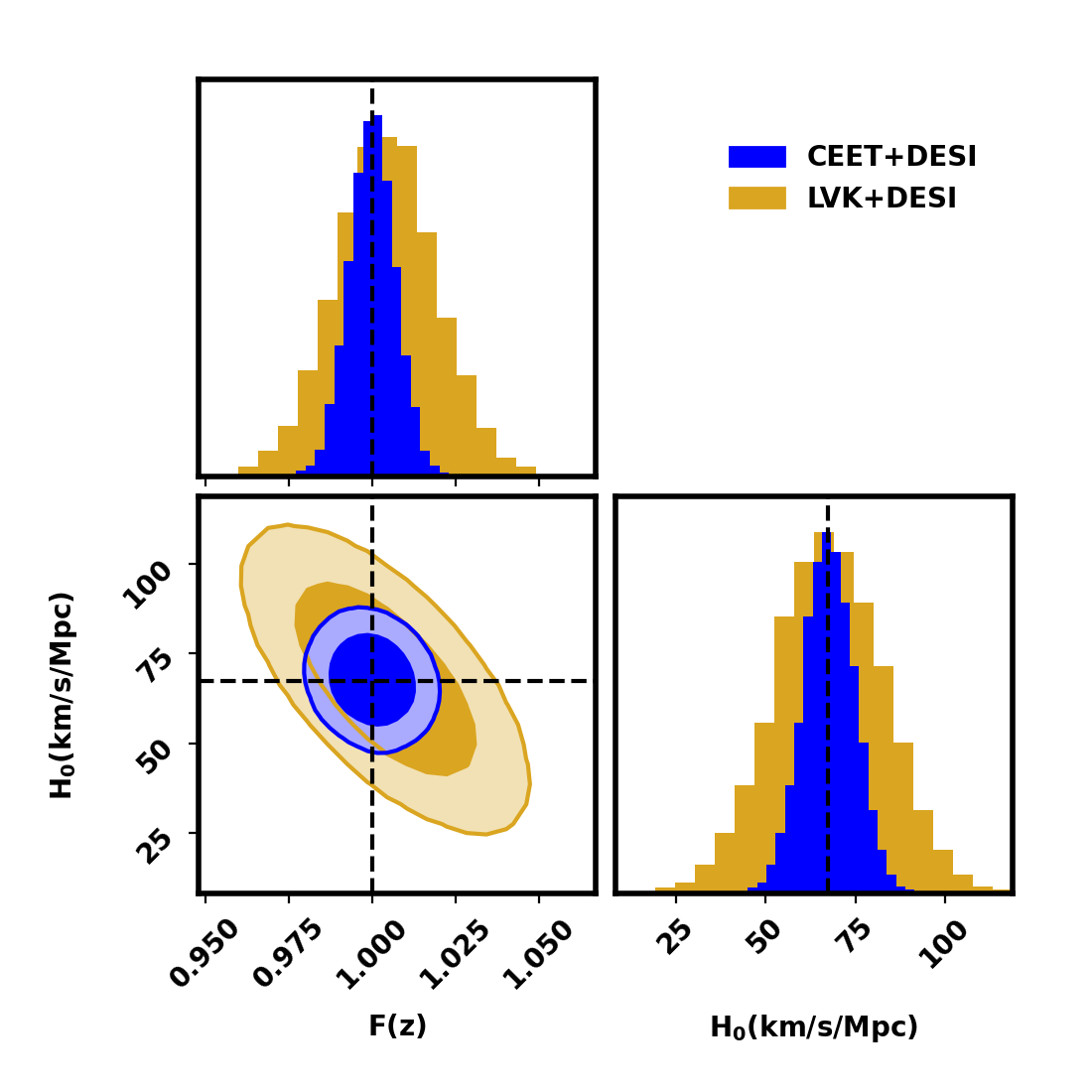}
\caption{This plot illustrates the best-combined measurements of F(z) and the Hubble constant ($\rm{H_0}$) for both the CEET and LVK systems using the DESI galaxy survey. These optimal measurements are obtained for an observation period of 5 years with a 75\% duty cycle and occur at a redshift of $\rm{z=0.425}$.}
\label{fig:combined}
\end{figure}

In our approach, we determine the EM luminosity distance from various length scales, including the BAO scale, sound horizon distance, and redshift. Figures \ref{fig:LVKviolinH0} and \ref{fig:CEETviolinH0}   provide a comprehensive comparison of $\rm{\mathcal{F}(z)}$ measurements under the scenario of varying  $\rm{H_0}$ for both the LVK and CEET sources respectively by combining with DESI and LSST. This comparative analysis highlights the impact of galaxy surveys on the precision of $\rm{\mathcal{F}(z)}$ measurements. Additionally, it underscores the potential of advanced detectors like CEET to deliver increasingly accurate cosmological insights, extending deep into higher redshifts. We also show in Figure \ref{fig:combined} the optimal combined measurements of $\rm{\mathcal{F}(z)}$ and the Hubble constant ($\rm{H_0}$) at $\rm{z=0.425}$ for both the CEET and LVK systems with the DESI galaxy survey. It shows that there exists a degeneracy between $\rm{H_0}$ and $\rm{\mathcal{F}(z)}$, as a result, the error on the individual parameters increases. Allowing \(\rm{H_0}\) to vary increases the errors in the measurements of \(\rm{\mathcal{F}(z)}\) by a factor of $\sim$ 2 compared to measurements with a fixed \(H_0\). The typical error on the measurements of $\rm{\mathcal{F}(z)}$ with a varying $\rm{H_0}$ scenario for LVK+DESI ranges around 19\% between $\rm{z=0.325}$ and $\rm{z=0.625}$, while for LVK+LSST it is around 23\%. Outside this range, the error increases significantly, as shown in Figure \ref{fig:LVKviolinH0}. Similarly, for CEET+DESI, the typical error is around 7.5\% between $\rm{z=0.275}$ and $\rm{z=1.2}$, and for CEET+LSST, it is around 9\%. Beyond these ranges, the error also increases significantly, as depicted in Figure \ref{fig:CEETviolinH0}.

\begin{figure}
\centering
\includegraphics[height=5.0cm, width=9cm]{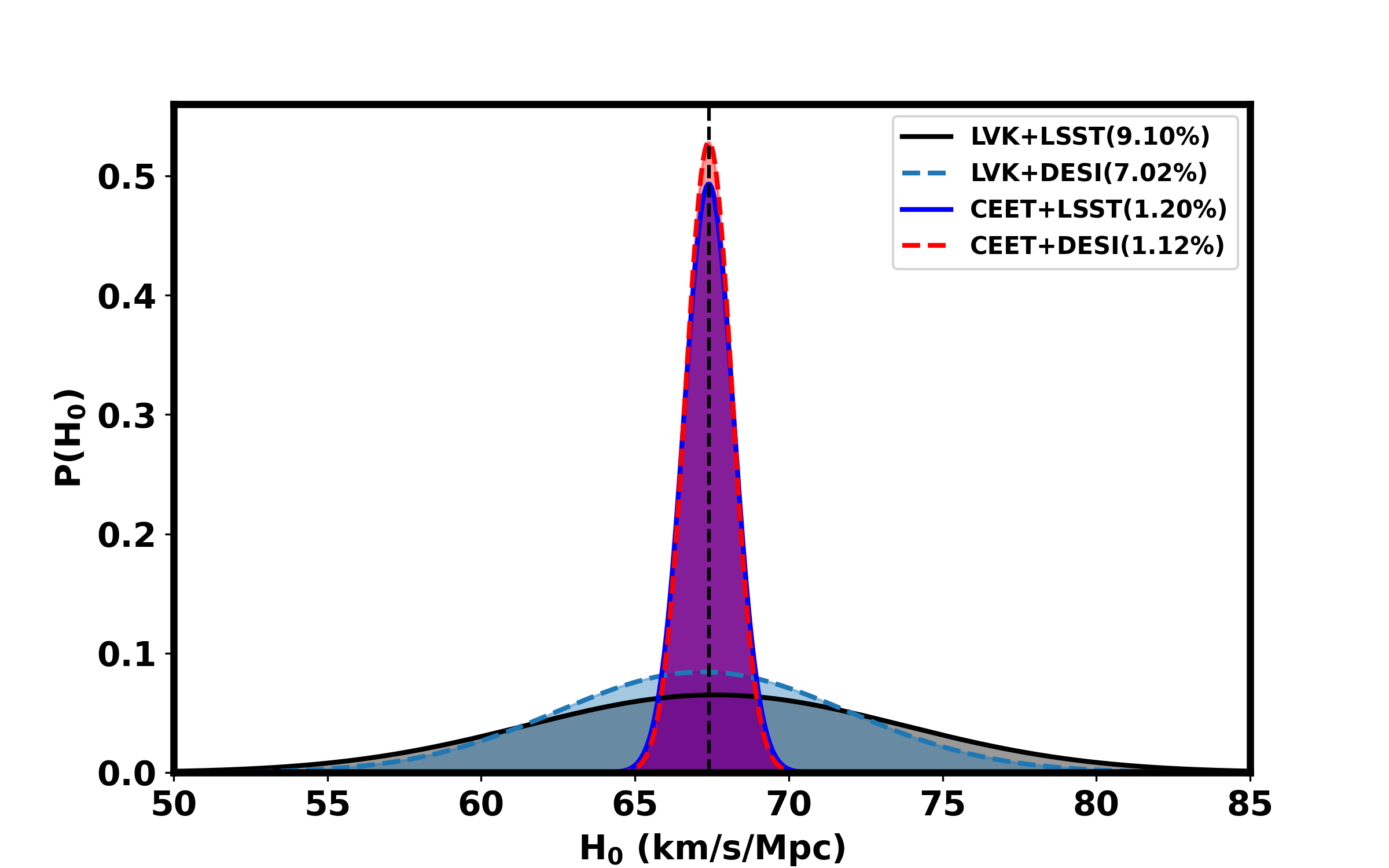}
\caption{The plot illustrates the posterior distribution of the Hubble constant, H0, for combined sources from the LVK GW detector with galaxy surveys DESI (LVK+DESI) and LSST (LVK+LSST), as well as the CE and ET detectors with DESI (CEET+DESI) and LSST (CEET+LSST) surveys. The analysis covers binary black hole mergers up to a redshift of $\rm{z=2.5}$ for CEET and up to $\rm{z=0.725}$ for LVK, considering five years of observational time with a 75\% duty cycle. The error associated with each combination is indicated in parentheses in the legend.}
\label{fig:H0}
\end{figure}

In Figure \ref{fig:H0}, we illustrate the posterior distribution of $\rm{H_0}$, obtained from the analysis of combined sources using the LVK detectors with galaxy surveys DESI (LVK+DESI) and LSST (LVK+LSST), as well as the CE and ET detectors with DESI (CEET+DESI) and LSST (CEET+LSST) surveys. This analysis encompasses binary black hole mergers up to a redshift of $\rm{z=2.5}$ for CEET and up to $\rm{z=0.725}$ for LVK, assuming five years of observational time with a 75\% duty cycle. The plot shows how the integration of different galaxy surveys with GW detections can influence the precision of the $\rm{H_0}$ measurement. The errors associated with each combination are provided in parentheses in the legend, highlighting the varying degrees of uncertainty depending on the survey and detector combination. Notably, the errors are impacted by the $\rm{\mathcal{F}(z)}$, and an inference considering only $\rm{H_0}$ would result in a smaller error, approximately a factor of 2.5 times less. This number can be compared with other dark siren analyses for BBHs \citep{Gair:2022zsa, Muttoni:2023prw}.

The primary source of uncertainty in \(\rm{\mathcal{F}(z)}\) measurements arises from redshift errors, which are estimated via cross-correlation between GW sources and galaxy surveys. This uncertainty decreases with increased sky coverage of the surveys, improving the precision of \(\rm{\mathcal{F}(z)}\) measurements. Accurate estimation of \(\rm{\mathcal{F}(z)}\) also depends on precise measurements of both the BAO scale and the GW luminosity distance, \(\rm{D_l^{\rm{GW}}(z)}\), as well as sky localization. LSST's photometric redshift errors, due to limited wavelength coverage and color-redshift degeneracies, and DESI's more precise spectroscopic redshifts, though still affected by minor errors, influence overall precision.Sky coverage is critical; LSST covers approximately 18,000 square degrees (\(\rm{f_{\text{sky}} \approx 0.436}\)) and DESI around 14,000 square degrees (\(\rm{f_{\text{sky}} \approx 0.339}\)). Expanding to full-sky coverage would significantly reduce redshift errors. Combining surveys with different sky areas enhances measurement precision by increasing effective sky coverage and the number of sources for cross-correlation, thereby improving cosmological parameter estimation. Additionally, GW luminosity distance measurements face uncertainties from waveform modeling, instrumental noise, inclination angle degeneracies, and weak lensing, while BAO measurements are affected by finite sample sizes and survey geometry. Statistical noise from the limited number and distribution of galaxies and GW sources further impacts inferred redshift accuracy.

\section{Conclusion \& Discussion}
\label{sec:conclusion}

In this study, we introduce a model-independent data-driven approach to investigate the frictional term, capturing any deviations from GR across redshifts using dark standard sirens. We present the reconstruction of the frictional term across redshifts for both the upcoming CEET and the currently operating LVK ground-based detectors using two major galaxy surveys LSST and DESI, focusing exclusively on dark sirens (GW sources without EM counterparts). Our approach involves reconstructing the frictional term as a function of redshift using a data-driven method. Specifically, we compare three distinct length scales with redshift: the luminosity distance from GW sources ($\rm{D_l^{GW}(z)}$), the angular scale of the BAO from galaxy surveys ($\rm{\theta_{BAO}(z)}$), and the sound horizon ($\rm{r_s}$) from CMB observations. 

In a previous study \citep{Afroz:2023ndy,Afroz:2024oui}, we demonstrated the feasibility of reconstructing the frictional term in a model-independent manner using bright standard sirens, relying on current and forthcoming ground and space-based detectors. Bright siren analyses, which require EM counterparts for accurate measurement of the frictional term, are inherently restricted to GW sources, where EM counterparts are more likely. However, the detectability of EM counterparts for all such GW events is not guaranteed due to observational constraints, and many of these counterparts, though existing, remain undetected. Moreover, the majority of detectable sources through current LVK detectors are stellar mass BBH, which are more numerous than GW sources with EM counterparts but cannot be utilized for measuring the frictional term with existing bright siren analysis methods due to the absence of EM counterparts. To overcome these limitations, the current study expands the scope by employing dark sirens. This approach not only addresses the challenges posed by the scarcity of EM counterparts but also significantly extends the applicability of our method to a broader array of GW sources. We demonstrate that by integrating BAO measurements from galaxies, sound horizon distances from the CMB power spectrum, redshift information from the cross-correlation of galaxy surveys with GW sources, and luminosity distance measurements from GW sources, one can perform a data-driven inference of the frictional term $\rm{\mathcal{F}(z)}$ and the Hubble constant $\rm{H_0}$. We have demonstrated the feasibility of measuring the frictional term using our ongoing detectors, such as LVK, across various redshifts. Additionally, we have explored the capabilities of upcoming ground-based detectors, specifically CEET, in quantifying the frictional term under various scenarios. We have demonstrated the feasibility of achieving precise measurements of the frictional term with an accuracy of approximately 3.6\% for CEET and 7.4\% for LVK at a redshift of $\rm{z=0.425}$ using this multi-messenger technique for DESI galaxy survey. Additionally, we have shown that the Hubble constant ($\rm{H_0}$) can be measured with a precision of about 1.1\% for CEET and 7\% for LVK, accounting for variations in the effective Planck mass over five years of observation time($\rm{T_{obs}}$) also for DESI galaxy survey. This analysis assumes a duty cycle of 75\%. This precision improves with observation time as $\rm{T_{obs}^{-1/2}}$.

This technique holds promise for dark standard sirens in the mHz range, detectable by future space-based detectors like LISA. To enhance the accuracy of these measurements, a high-quality spectroscopic survey is crucial. Such a survey would provide precise spectroscopic redshifts for galaxies, reducing the impact of photometric redshift errors and improving the overall accuracy of our cosmological distance measurements. Moreover, a full-sky survey and an increased number of galaxy sources are essential to further refine our measurements of $\rm{\mathcal{F}(z)}$ and $\rm{H_0}$, along with advancements in GW source detection and analysis techniques. In the future, this technique will be applied on the GW sources data from LVK to test GR along with cosmology in synergy with DESI and LSST.

\section*{Acknowledgements}
The authors express their gratitude to Ulyana Dupletsa for reviewing the manuscript and providing useful comments as a part of the LIGO publication policy. This work is part of the \texttt{⟨data|theory⟩ Universe-Lab}, supported by TIFR and the Department of Atomic Energy, Government of India. The authors express gratitude to the computer cluster of \texttt{⟨data|theory⟩ Universe-Lab} and the TIFR computer center HPC facility for computing resources. Special thanks to the LIGO-Virgo-KAGRA Scientific Collaboration for providing noise curves. LIGO, funded by the U.S. National Science Foundation (NSF), and Virgo, supported by the French CNRS, Italian INFN, and Dutch Nikhef, along with contributions from Polish and Hungarian institutes. This collaborative effort is backed by the NSF’s LIGO Laboratory, a major facility fully funded by the National Science Foundation. The research leverages data and software from the Gravitational Wave Open Science Center, a service provided by LIGO Laboratory, the LIGO Scientific Collaboration, Virgo Collaboration, and KAGRA. Advanced LIGO's construction and operation receive support from STFC of the UK, Max-Planck Society (MPS), and the State of Niedersachsen/Germany, with additional backing from the Australian Research Council. Virgo, affiliated with the European Gravitational Observatory (EGO), secures funding through contributions from various European institutions. Meanwhile, KAGRA's construction and operation are funded by MEXT, JSPS, NRF, MSIT, AS, and MoST. This material is based upon work supported by NSF’s LIGO Laboratory which is a major facility fully funded by the National Science Foundation.

This work has made use of CosmoHub, which has been developed by the Port d'Informació Científica (PIC), maintained through a collaboration of the Institut de Física d'Altes Energies (IFAE) and the Centro de Investigaciones Energéticas, Medioambientales y Tecnológicas (CIEMAT) and the Institute of Space Sciences (CSIC \& IEEC). CosmoHub was partially funded by the 'Plan Estatal de Investigación Científica y Técnica y de Innovación' program of the Spanish government, has been supported by the call for grants for Scientific and Technical Equipment 2021 of the State Program for Knowledge Generation and Scientific and Technological Strengthening of the R+D+i System, financed by MCIN/AEI/10.13039/501100011033 and the EU NextGeneration/PRTR (Hadoop Cluster for the comprehensive management of massive scientific data, reference EQC2021-007479-P) and by MICIIN with funding from European Union NextGenerationEU(PRTR-C17.I1) and by Generalitat de Catalunya.

We acknowledge the use
of the following packages in this work: Astropy \citep{robitaille2013astropy, price2018astropy}, Bilby \citep{ashton2019bilby}, Pandas \citep{mckinney2011pandas}, NumPy \citep{harris2020array}, Seaborn \citep{bisong2019matplotlib}, Scipy \citep{virtanen2020scipy}, Dynesty \citep{speagle2020dynesty}, emcee \citep{foreman2013emcee}, Nbodykit \citep{Hand:2017pqn} and Matplotlib \citep{Hunter:2007}.

\section*{Data Availability}
The corresponding author will provide the underlying data for this article upon request.

\bibliographystyle{mnras}
\bibliography{references} 
\label{lastpage}

\bsp
\end{document}